\documentclass[fleqn,usenatbib]{mnras}

\usepackage{newtxtext,newtxmath}
\usepackage[T1]{fontenc}
\usepackage{ae,aecompl}

\usepackage{graphicx}
\usepackage{bm}		
\usepackage[flushleft]{threeparttable}
\usepackage{float}
\usepackage[caption = false]{subfig}
\usepackage{hyperref}
\usepackage[normalem]{ulem}
\AtBeginShipout{%
  \ifnum\value{page}>1 %
    \typeout{* Additional boxing of page `\thepage'}%
    \setbox\AtBeginShipoutBox=\hbox{\copy\AtBeginShipoutBox}%
  \fi}

\usepackage{multicol}        

\usepackage{pdflscape}
\usepackage{adjustbox}
\usepackage{threeparttable}

\title[X-ray spectral variability in SWIFT J2127.4+5654]{The nature of X-ray spectral variability in SWIFT\,J2127.4+5654}

\author[E. S. Kammoun and I. E. Papadakis]{E. S. Kammoun,$^{1}$\thanks{E-mail: \href{mailto:ekammoun@sissa.it}{ekammoun@sissa.it}}
and I. E. Papadakis$^{2,3}$ 
\\
$^{1}$SISSA, via Bonomea 265, I-34135 Trieste, Italy\\
$^{2}$Department of Physics and Institute of Theoretical and Computational Physics, University of Crete, 71003, Heraklion, Greece\\
$^{3}$IESL, Foundation of Research and Technology, 71110 Heraklion, Greece\\
}
\date{Accepted XXX. Received YYY; in original form ZZZ}

\pubyear{2017}

\begin{document}
\label{firstpage}
\pagerange{\pageref{firstpage}--\pageref{lastpage}}

\maketitle

\begin{abstract}
We study the flux-flux plots (FFPs) of the Seyfert galaxy SWIFT~J2127.4+5654 which was observed simultaneously by {\it XMM-Newton} and {\it NuSTAR}. The 0.7--2 keV FFPs show a non-linear behaviour, while they are well fitted by a straight line, in the $\sim 2-40$ keV range. Without any additional modelling, this result strongly suggests that neither absorption nor spectral slope variations can contribute significantly to the observed variability above 2 keV (on time scales $\lesssim$5 days, at least). The FFP method is capable of separating the variable and the constant X-ray spectral components in this source. We found that the low-, average- and high-flux spectra of the variable components are consistent with a power law varying in normalisation only plus a relativistic reflection component varying simultaneously with the continuum. At low energies, we identify an additional variable component, consistent with a neutral absorber with a variable covering fraction. We also detect two spectral components whose fluxes remain constant over the duration of the observations. These components can be well described by a blackbody emission of $kT_{\rm BB}\sim 0.18\, \rm keV$, dominating at soft X-rays, and a reflection component from neutral material. The spectral model resulting from the FFP analysis can fit well the energy spectra of the source. We suggest that a combined study of FFPs and energy spectra in the case of X-ray bright and variable Seyferts can identify, unambiguously, the various spectral components, and can lift the degeneracy that is often observed within different models fitting equally well time-averaged spectra of these objects.
\end{abstract}

\begin{keywords}
galaxies: active - galaxies: individual: SWIFT\,J2127.4+5654 -- galaxies: nuclei -- galaxies: Seyfert-- X-rays: galaxies.
\end{keywords}


\section{Introduction}
\label{sec:intro}

It is commonly thought that active galactic nuclei (AGN) host a supermassive black hole (BH), at their center, that is accreting matter through a geometrically thin optically thick disc \citep{Shak73}. The X-ray emission in non-jetted AGN is widely accepted to be due to Compton upscattering of ultraviolet (UV)/soft X-ray disc photons off hot electron \citep[$\sim 10^9$\,K; e.g.][]{Shap76, Haa93}, usually referred to as the `X-ray corona'. The high-amplitude, short-timescale, X-ray variability that is observed in these sources suggests that the emitting region is compact and located in the innermost region of AGN.

In this work, we use the flux-flux plot (FFP) method to study the X-ray spectral variability of SWIFT\,J2127.4+5654. This method was first suggested by \cite{Chu01} in order to study the X-ray variability of the BH binary Cygnus X-1. It was then applied to bright Seyferts by \cite{Tay03}, and has been used many times since then. We have recently applied this method to the {\it XMM-Newton} observations of IRAS~13224--3809 \citep{Kammoun15} and to the simultaneous {\it XMM-Newton} and {\it NuSTAR} observations of MCG--6-30-15 \citep[][hereafter KP17]{Kammoun17}. In the last work, we emphasized the advantages of the method when {\it NuSTAR} observations are available. For that reason, we chose to study SWIFT\,J2127.4+5654 which is an X--ray bright and variable AGN, and was observed simultaneously by {\it XMM-Newton} and {\it NuSTAR} over a period of $\sim 5$ days in November, 2012.

SWIFT\,J2127.4+5654 \citep[a.k.a IGR J21277+5656, $z=0.0144$, $M_{\rm BH} = 1.5\times 10^7\, M_\odot$;][]{Malizia08} is a Seyfert\,1 galaxy that was first detected in X-rays by {\it Swift}/BAT \citep{Tueller05}. It is also detected in the radio band with a 20\,cm flux of 6.4 mJy, without showing any visible elongation in the 1.4 GHz NRAO VLA Sky Survey \citep[NVSS;][]{Condon98}. This source was observed for 92~ks with {\it Suzaku}-XIS in 2007 \citep{Miniutti09}. These authors detected a broad Fe K$\alpha$ line and estimated a BH spin of $\sim 0.6$. This result was later confirmed by \cite{Sanfrutos13} who analysed the $\sim 130$~ks observation with {\it XMM-Newton} in 2010. They also detected significant X--ray spectral variability which they suggested was due to an eclipsing event by a cloud of $N_{\rm H} = 2 \times 10^{22}\,\rm cm^{-2}$, with the covering fraction varying between 0 and 0.43. \cite{Marinucci14Swift} have presented the results from the spectral study of the the 2012 simultaneous {\it XMM-Newton} and {\it NuSTAR} observations of SWIFT\,J2127.4+5654, while \cite{Kara15} used the same data sets and estimated the reverberation time lags in the source. 

In Section \ref{sec:obsred} we present the observations and the data reduction. In \ref{sec:FFA} we present the FFP analysis and we interpret the results in Section \ref{sec:results}. In Section \ref{sec:energy_spectra} we present the modelling of the low- and high- flux
observed energy spectra of the source, based on the FFP analysis results. Finally, we discuss the implications of our results in Section \ref{sec:conclusion}.

\section{Observations and data reduction}
\label{sec:obsred}

\subsection{\it XMM-Newton}
\label{subsec:XMMdata}

The {\it XMM-Newton} satellite \citep{Jans01} observed SWIFT\,J2127.4+5654 during three consecutive revolutions (Obs. IDs 0693781701, 0693781801, and 0693781901), simultaneously with {\it NuSTAR} \citep{Har13}, starting on 2012 November 4 (PI: G. Matt). The data are available in the {\it XMM-Newton} Science Archive\footnote{\url{http://nxsa.esac.esa.int/nxsa-web}} (XSA). We considered data only from the EPIC-pn camera \citep{Stru01}, that was operating in small window/medium filter imaging mode. 
We reduced the data using the {\it XMM-Newton} Science Analysis System ({\tt SAS}\,v15.0.1) and the latest calibration files. The data were cleaned for strong background flares and were selected using the criterion PATTERN\,$\leq$\,4. Source light curves were extracted from a circle of radius 40\arcsec, while the background light curves were extracted from an off-source circular region of radius 50\arcsec. We checked for pileup and we found it to be negligible in all observations. Background-subtracted light curves were produced using the {\tt SAS} task {\tt EPICLCCORR}.

\subsection{\it NuSTAR}
\label{subsec:Nustardata}

Swift\,J2127.4+5654  was observed by {\it NuSTAR} with its two co-aligned telescopes with corresponding Focal Plane Modules\,A (FPMA) and B (FPMB) starting on 2012 November 4 (Obs. IDs 60001110002, 60001110003, 60001110005, 60001110007). We reduced the {\it NuSTAR} data following the standard pipeline in the {\it NuSTAR} Data Analysis Software (NuSTARDAS\,v1.6.0). We used the instrumental responses from the latest calibration files available in the {\it NuSTAR} calibration database (CALDB). The unfiltered event files were cleaned with the standard depth correction, which reduces the internal background at high energies, and we excluded South Atlantic Anomaly passages from our analysis. The source and background light curves were extracted from circular regions of radii 1\farcm5 and 3\arcmin, respectively, for both FPMA and FPMB, using the HEASoft task {\tt NUPRODUCT}, and requiring an exposure fraction larger than 50\,\%. We checked that the background-subtracted light  curves of the two {\it NuSTAR} modules were consistent with each other as follows. We divided the FPMA over the FPMB light curves (binned at $\Delta t = 1\,{\rm ks}$), in all the energy bands we consider in this work (see next Section), and we fitted the ratio as a function of time with a constant, $C$. The fit was acceptable in all cases, indicating that the FPMA and FPMB light curves are consistent. Given this result, we added the FPMA and FPMB light curves using the {\tt FTOOLS} \citep{Ftools} command {\tt LCMATH}, in order to increase the signal-to-noise of the {\it NuSTAR} light curves. 

\begin{figure*}
\centering
\includegraphics[width =1.0\textwidth]{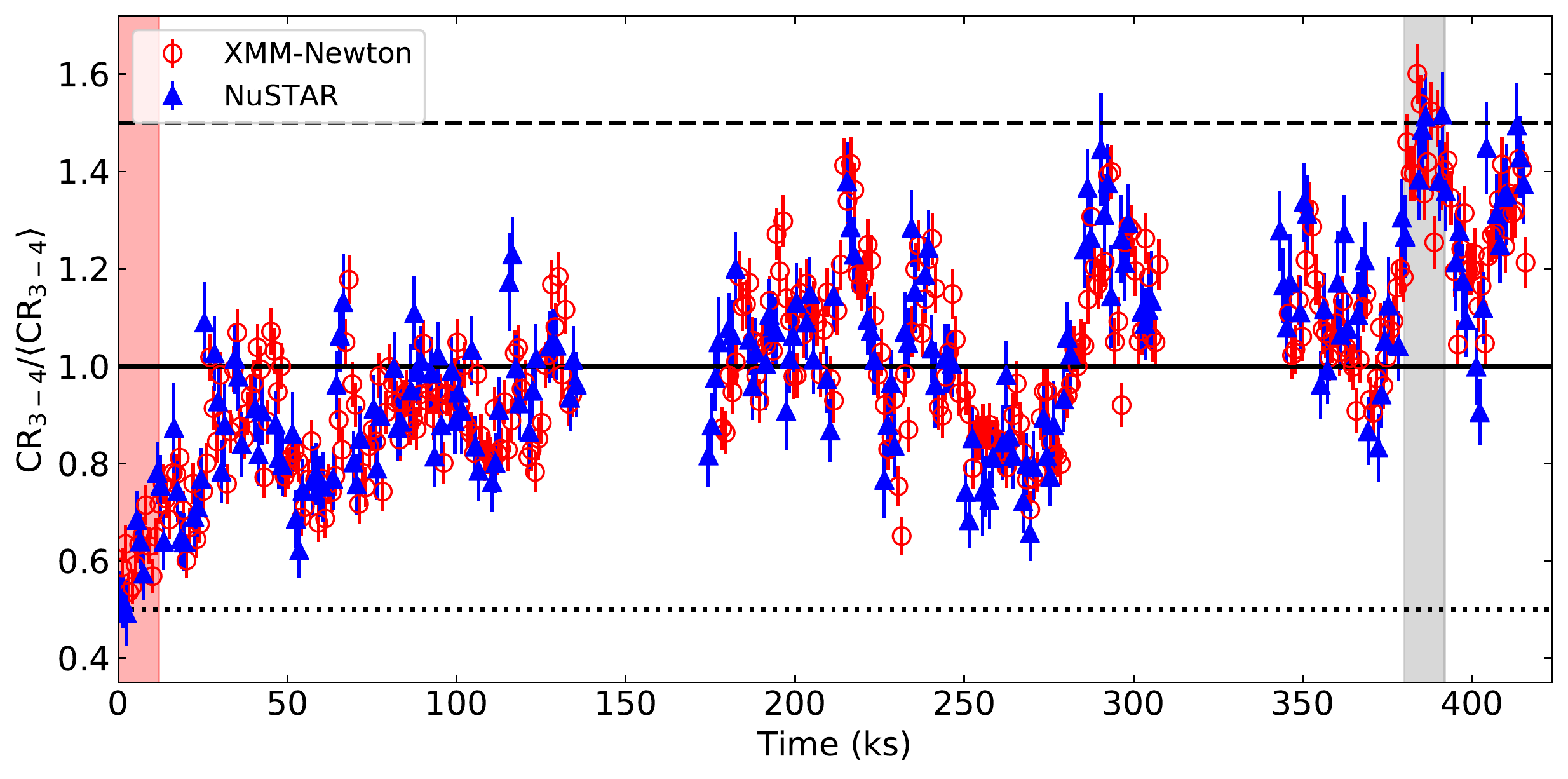}
\caption{The 1\,ks binned, normalized, 3--4 keV band EPIC-pn, and {\it NuSTAR} light curves (open red circles, filled blue triangles, respectively). Time is measured from the start of {\it NuSTAR} observations. The dotted, solid and dashed black lines indicate the lowest, mean and highest source flux (see Section \,\ref{subsec:high-variable}). The red and grey shaded areas indicate the LF and HF  intervals (see Section \,\ref{sec:energy_spectra} for details).}
\label{fig:lightcurve}
\end{figure*}

Figure\,\ref{fig:lightcurve} shows the {\it XMM-Newton} and {\it NuSTAR} light curves in the 3--4\,keV band (chosen to be the reference band in the FFP analysis; see below), normalized to the mean count rate. We plot data only when both satellites were observing the source. We considered data from these periods only, by merging the good time intervals tables of the two satellites using the {\tt FTOOLS} command {\tt MGTIME}. This figure shows the variability range of the source (the max-to-min flux ratio is $\sim 3.3$), and the consistency between the instruments. The log of the observations is presented in Table\,\ref{table:logTable}.

\begin{table}
\centering
\caption{Net exposure time and the net count rate in the 3--4 keV band for the various observations and instruments considered in this work.}
\begin{adjustbox}{max width=1\linewidth}
\begin{tabular}{lcccc}
\hline
Obs	&	Exp. Time (ks)	&						\multicolumn{3}{c}{$ {\rm net\, CR_{3-4} \,(Count\,s^{-1}) } $ }						\\[0.1cm]  	\cline{3-5} \\[-0.1cm]
	&	EPIC-pn/FPMA,B	&		EPIC-pn		&		FPMA		&		FPMB		\\  \hline 	
1	&	94.5/77.7	&	0.524	$ \pm $	0.002	&	0.085	$ \pm $	0.002	&	0.080	$ \pm $	0.002	\\[0.1cm]   	
2	&	91.9/74.6	&	0.632	$ \pm $	0.003	&	0.106	$ \pm $	0.001	&	0.104	$ \pm $	0.001	\\[0.1cm]   	
3	&	50.1/42.1	&	0.699	$ \pm $	0.004	&	0.121	$ \pm $	0.002	&	0.118	$ \pm $	0.002	\\[0.1cm]   \hline	
\end{tabular}
\end{adjustbox}
\label{table:logTable}
\end{table}

\section{Flux-flux analysis}
\label{sec:FFA}
\subsection{Choice of the energy bands and time bin size}

\begin{figure}
\centering
\includegraphics[width =1.0\linewidth]{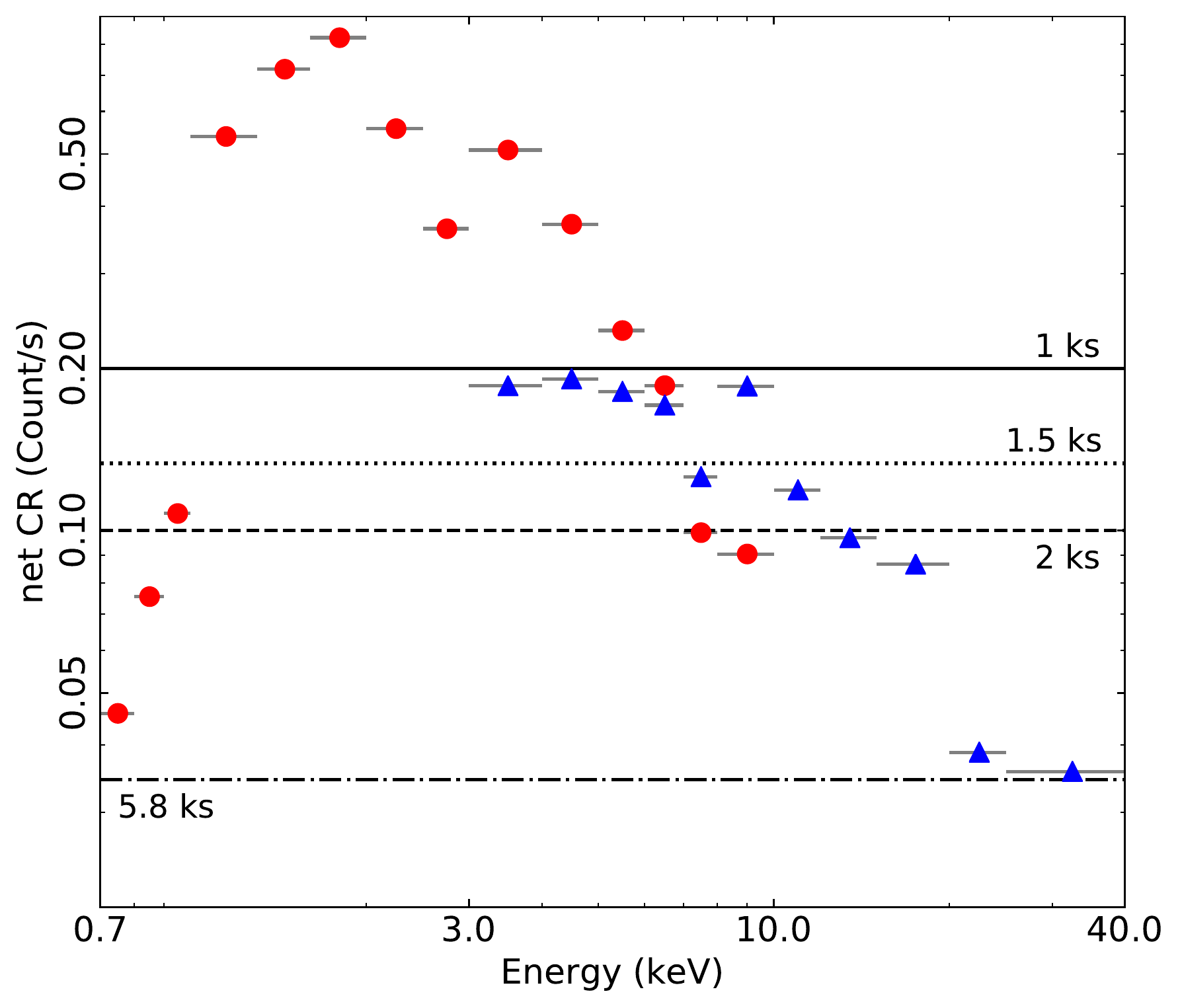}
\caption{The observed mean count rate in the various energy bands for the {\it XMM-Newton} (red circles) and {\it NuSTAR} (blue triangles), extracted from the first observation. The horizontal lines indicate the 200 Count limit in the case of 1\,ks, 1.5\,ks, 2\,ks and 5.8\,ks bins.}
\label{fig:netCR}
\end{figure}

KP17 showed that an average of $\sim 200$ counts in both energy bands is needed, in order to avoid the effect of Poisson noise in the FFPs. Hence, we chose our energy bands and bins accordingly. The source is heavily absorbed below $\sim 0.7$\,keV, due to high Galactic absorption in the line of sight \citep[$N_{\rm H}= 7.65 \times 10^{21}\,{\rm cm^{-2}}$;][]{Kal05}. For that reason, we constructed FFPs at energies larger than 0.7~keV, and we considered 19 energy bands up to 40~keV (at higher energies the source count rate is too low). Fig.\,\ref{fig:netCR} shows the observed mean count rate of the {\it XMM-Newton} and {\it NuSTAR} light curves in these bands, using data from the first observation only (when the source flux was the lowest). The horizontal lines in the same figure indicate the 200-count limit for various bin sizes (5.8~ks is the orbital period of {\it NuSTAR}). The numbers in parenthesis in the first column of Tables\,\ref{table:bestfitlinearXMM}-\ref{table:bestfitPLcXMM} are the bin size we chose so that the average counts would be above 200 in each light curve. Similarly to KP17, we chose the reference band to be the 3--4~keV band, as it is common to both instruments and (most probably) is representative of the primary component. The FFPs are plotted in Figs.\,\ref{figapp:commFFPs}-\ref{figapp:lowEFFPs}.

\subsection{The high-energy flux-flux plots}
\label{subsec:highE-FFP}

\begin{figure}
\centering
\includegraphics[width = 1.\linewidth]{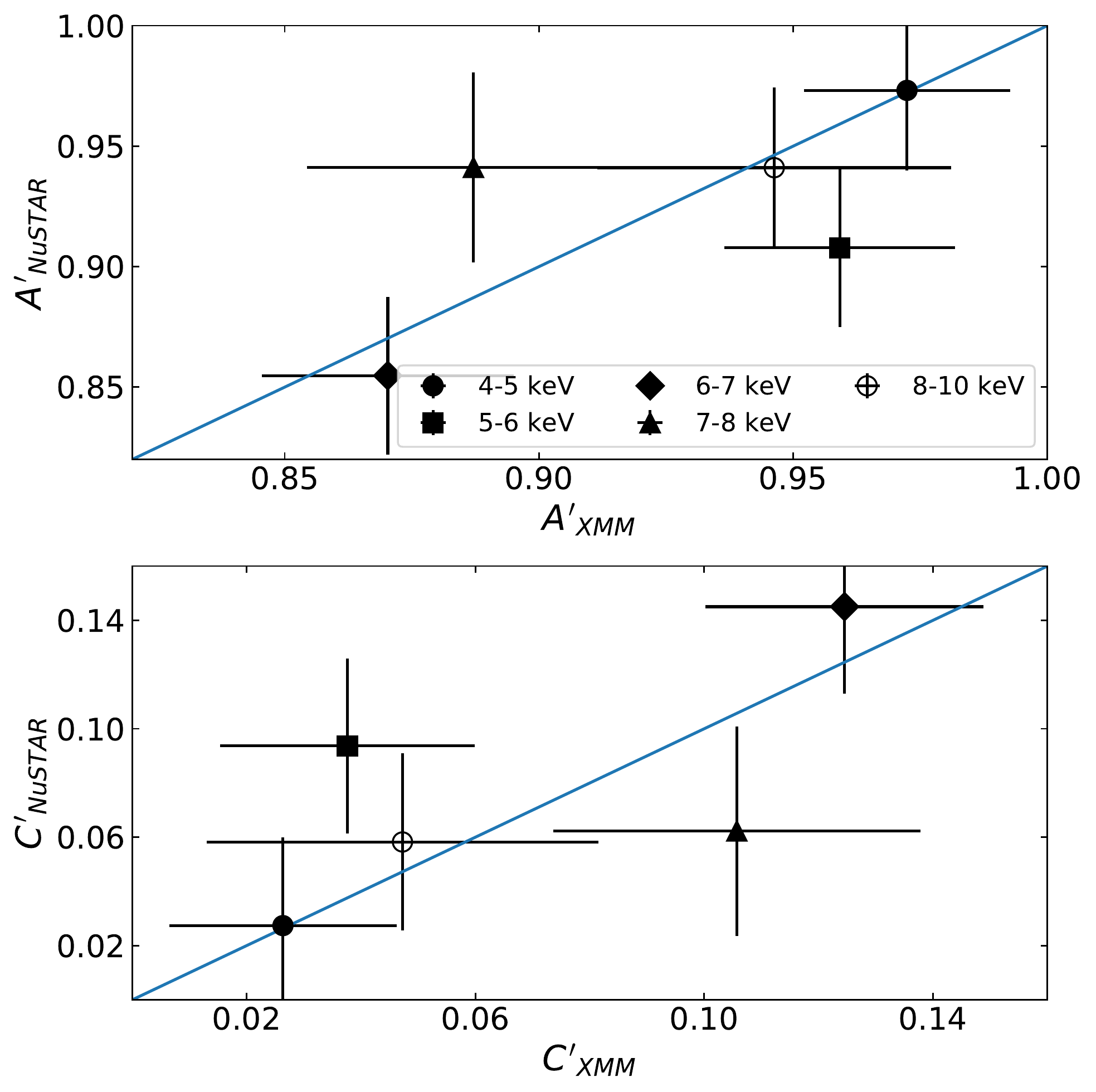}
\caption{Plot of the {\it NuSTAR} versus the {\it XMM-Newton} normalized $A_{\rm L}$ (top panel) and $C_{\rm L}$ (bottom panel) best-fit  values. The solid lines indicate the one-to-one relationship.}
\label{fig:XNcomp}
\end{figure}

We fitted the combined FFPs above 3\,keV (using data from all observations) with a linear model of the form, 
\begin{equation}
 y = A_{\rm L} x + C_{\rm L},
 \label{eq:linear}
\end{equation}
\noindent
($x$ in this, and all equations hereafter, represents the count rate in the reference band). We used the {\tt MPFFITEXY} routine \citep{Williams10} which takes into account the errors on both $x$ and $y$ variables. Tables \ref{table:bestfitlinearXMM} and \ref{table:bestfitlinearNustar} in Appendix\,\ref{app:tables} list the best-fit results. Best-fit models are plotted as solid lines in Figs.\,\ref{figapp:commFFPs} and \ref{figapp:nustarFFPs}. We also fitted the FFPs from the individual observations separately. The best-fit parameters were consistent between the three observations and with the best-fit values obtained by fitting the combined FFPs. In order to check the consistency between the best-fit parameters obtained from the {\it XMM-Newton} and {\it NuSTAR} FFPs, we show in Fig.\,\ref{fig:XNcomp} the normalized, best-fit, $A'_{\rm L}$ and $C'_{\rm L}$ values (see eq. 3 in KP17) from the {\it NuSTAR} FFPs versus the respective {\it XMM-Newton} values. The agreement with the one-to-one line (indicated by the diagonal, solid line in the two panels) demonstrates the full consistency between the two instruments.
 
A straight line fits well ($p$-value $>0.01$) many, but not all, FFPs. The last columns of Tables\,\ref{table:bestfitlinearXMM} and \ref{table:bestfitlinearNustar} list the root mean square deviation, $\sigma_{\rm rms}$, of the data from the model, calculated as in \cite{Kammoun15}. The average data-to-model deviations are of the order of $\sim 5$\% and $\sim 7$\% for {\it XMM-Newton} and {\it NuSTAR}, respectively. The best-fit residuals shown in Figs.\,\ref{figapp:commFFPs} and \ref{figapp:nustarFFPs} show a random scatter around zero, with no indication of any systematic trends. We therefore conclude that a straight line fits well the high-energy FFPs. Only a $\sim 5-7$\% of the total scatter around the mean is due to fast and random variations which are uncorrelated between the various energy bands. 

The linear behaviour of the FFPs at high energies is indicative of a power-law (PL) primary emission that varies in normalization with constant slope $\Gamma$, as discussed in KP17. In this case, the slope of the FFPs, $A_{\rm L}$, should be indicative of $\Gamma$. We fitted the best-fit $A_{\rm L}(E)$'s as a function of energy following KP17 (see their Section 3.3). We found that the observed {\it XMM-Newton} and {\it NuSTAR} $A_{\rm L}(E)$'s can be well reproduced by a variable PL component with $\Gamma_{\rm X} = 1.84 \pm 0.02$ ($\chi^2/\rm dof = 6.5/4$) and  $\Gamma_{\rm N} = 1.98 \pm 0.02$ ($\chi^2/\rm dof = 16.8/9$), respectively.

\subsection{The low-energy flux-flux plots}
\label{subsec:softEFFP}

The magenta dashed lines in Fig.\,\ref{figapp:lowEFFPs} show  the expected FFPs at energies below 3~keV, assuming a PL spectrum with $\Gamma_{\rm X} = 1.84$ that varies only in normalization (as suggested by the high energy FFPs). Clearly, the normalization of the FFPs below $\sim 2$~keV is significantly larger than predicted. This is indicative of an additional spectral component at low energies. The shape of the low energy band FFPs does not appear to be consistent with a straight line either. Indeed a linear model does not provide a statistically accepted fit neither to the combined FFPs nor to the FFPs from the individual observations at low energies, and the best-fit results do not agree with each other. The best-fit residuals showed a flattening at low count rates. We therefore re-fitted the combined FFPs assuming a power law plus a constant (PLc) model of the form,
\begin{equation}
 y = A_{\rm PLc} x^\beta + C_{\rm PLc}.
 \label{eq:plc}
\end{equation}
\noindent
We used the {\tt MPFIT}\footnote{\url{http://code.google.com/p/astrolibpy/source/browse/trunk/}} package \citep{Mark09}, taking into account the errors on the $y$-axis only. The best-fit parameters are reported in Table\,\ref{table:bestfitPLcXMM}. The PLc model is not statistically accepted. However, the residuals are randomly scattered around the zero without showing any systematic trend. The last column in Table\,\ref{table:bestfitPLcXMM} represents $\sigma_{\rm rms}$ that is $\sim 6$\% on average, comparable to the values that we obtain for the high-energy FFPs. This indicates that a PLc model can account for most of the variability in the soft band.

\section{Results}
\label{sec:results}
{In the following, we use the FFPs results presented in Tables \ref{table:bestfitlinearXMM}-\ref{table:bestfitPLcXMM}, in order to create the spectra of the variable and constant components for both {\it XMM-Newton} and {\it NuSTAR} (see later for more details). We first created an text file with as many entries as the channels in the original response files of EPIC-pn and FPMA. We assign a value which corresponds to the best-fit FFP parameter value to each channel (the same value, in count/s, is given to the various channels in the broad energy bands that we used to construct the FFPs). Then we used the {\tt ASCII2PHA} tool to create {\tt pha} files, and we grouped the spectra according to the energy bins that we consider in the FFPs, using the {\tt GRPPHA} task. In the case of {\it NuSTAR}, the use of the FPMB response matrices gives consistent results with the ones using the FPMA response matrices. The spectra are fitted using XSPEC v12.9s \citep{Arn96}.}

\subsection{The variable spectral components at high energies}
\label{subsec:high-variable}

\begin{figure}
\centering
\includegraphics[width=\linewidth]{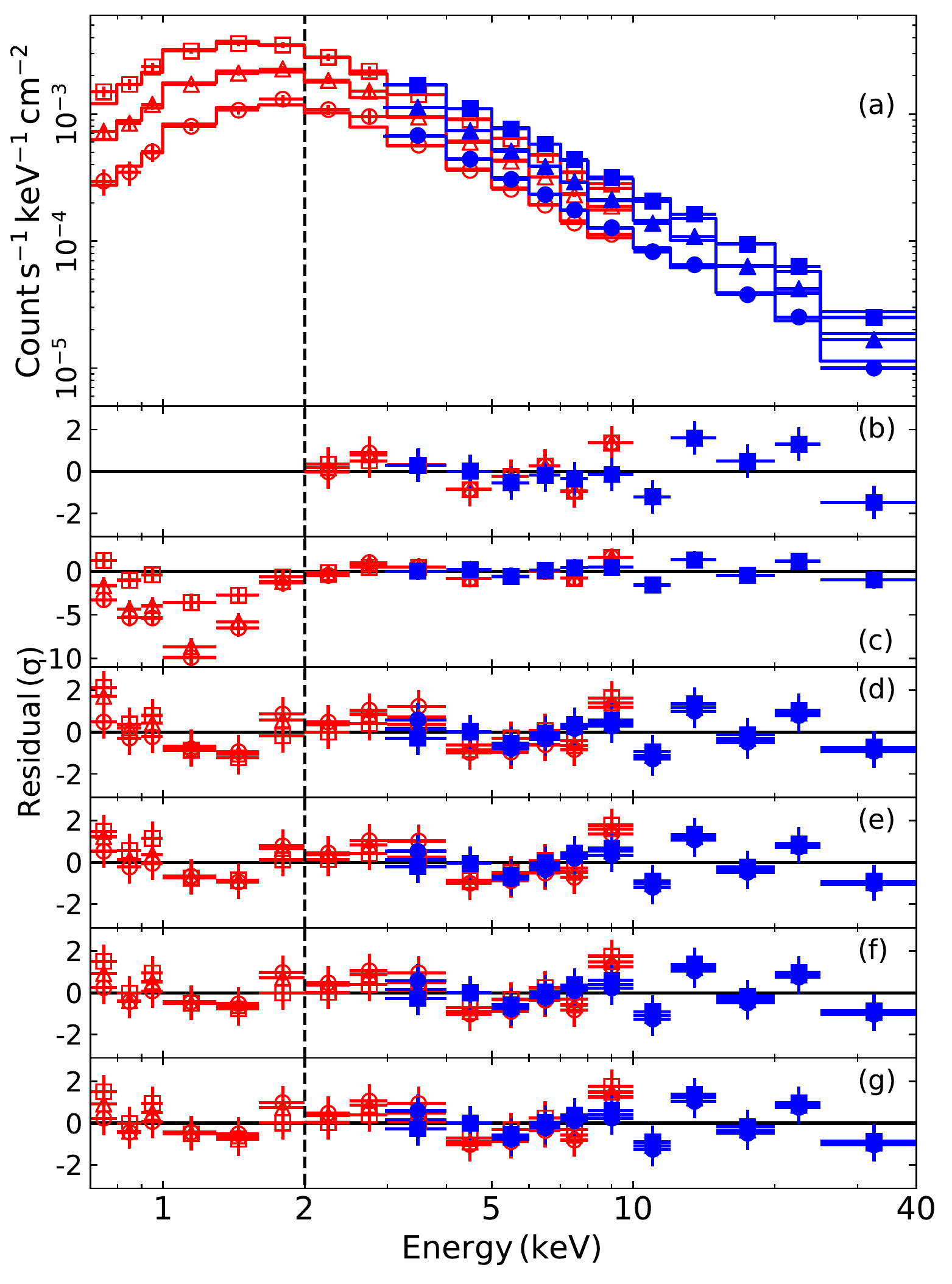}
\caption{Panel a: The {\it XMM-Newton} and {\it NuSTAR} spectra of the variable components (open and filled points, respectively). The solid lines indicate the Model C$'$ (see Table\,\ref{table:bestfitRelxill}) best-fit to the low-, average- and high flux spectra (circles, triangles, and squares, respectively). Panel (b): Best-fit residuals in the case of a PL fit above 2 keV. Panel (c): Best-fit residuals in the case of a {\tt RELXILL} (plus Galactic absorption only) fit above 2~keV (the residuals are shown at lower energies as well). Panels (d-g): Model A, B, C and C$'$ best-fit residuals (see Section \,\ref{subsec:high-variable}-\ref{subsec:low-variable} for details).}
\label{fig:variableSpec}
\end{figure}

\begin{table}
\centering
\caption{Best-fit parameters obtained by fitting the spectra of the variable component (see Section \ref{subsec:high-variable}-\ref{subsec:low-variable}). The column densities are in units of $10^{22} \rm cm^{-2}$. {The normalisations correspond to the mean-flux spectra ($N_{\rm m}$, in units of $10^{-4}~ \rm photon\,keV^{-1}\,cm^{-2}\,s^{-1}$). The normalisations of low- and high-flux spectra are consistent with 0.5 and 1.5 times $N_{\rm m}$.} The lower/upper limits correspond to 3$\sigma$ limits.}
\begin{adjustbox}{max width=1.\linewidth}
\begin{threeparttable}
\begin{tabular}{lccccc}
\hline	  
	&		$E>2$\,keV								&				model A						&				model B						&			model C							&		model C$'$																		\\[0.2cm]   \hline  
\multicolumn{6}{c}{\tt RELXILL}																																																													\\[0.2cm]   
$\Gamma_{\rm X}$	&	$	2.05	_{-	0.02	}	^{+	0.01	}	$	&	$	2.06	\pm	0.02					$	&	$	2.08	_{-	0.01	}	^{+	0.02	}	$	&	$	2.09	\pm	0.02					$	&	$	2.1	\pm	0.02					$											\\[0.2cm]   
$\Gamma_{\rm N}$	&	$	2.16	_{-	0.07	}	^{+	0.01	}	$	&	$	2.16	\pm	0.03					$	&	$	2.19	\pm	0.02					$	&	$	2.19	\pm	0.01					$	&	$	2.2	\pm	0.02					$											\\[0.2cm]   
$i (^\circ)$	&	$	28	_{	-19	}	^{+	12	}	$	&	$	35	_{-	16	}	^{+	11	}	$	&	$	36	_{-	16	}	^{+	10	}	$	&	$	30	_{-	17	}	^{+	12	}	$	&	$	30	_{-	18	}	^{+	12	}	$											\\[0.2cm]   
$\log \xi_{\rm d}$	&	$	<2.44							$	&	$	1.37	\pm	0.33					$	&	$	1.43	_{-	0.37	}	^{+	0.24	}	$	&	$	1.44	_{-	0.46	}	^{+	0.19	}	$	&	$	1.4	\pm	0.2					$											\\[0.2cm]   
$A_{\rm Fe}$\,(solar)	&	$	<0.86							$	&	$	< 0.77							$	&	$	 < 0.78							$	&	$	<0.78							$	&	$	< 0.78							$											\\[0.2cm]   
$E_{\rm cut}$\,(keV)	&	$	> 95							$	&	$	>78							$	&	$	>83							$	&	$	>92							$	&	$	>125							$											\\[0.2cm]  
$N_{\rm m,\, X}$	&	$	1.86	\pm	0.1					$	&	$	1.79	\pm	0.2					$	&	$	1.92	\pm	0.1					$	&	$	1.96	_{-0.15}	^{+0.09}					$	&	$	1.96	_{-0.15}	^{+0.09}					$\\[0.2cm]
$N_{\rm m,\,N}$	&	$	2.62	\pm	0.15					$	&	$	2.69	\pm	0.2					$	&	$	2.88	\pm	0.1					$	&	$	2.89	\pm	0.1					$	&	$	2.89	\pm	0.1					$
\\[0.2cm]\hline  
\multicolumn{6}{c}{\tt ZXIPCF}																																																													\\[0.2cm]    
$N_{\rm H,l}$	&	$	-							$	&	$	1.08	_{-	0.15	}	^{+	0.09	}	$	&	$	1.07	_{-	0.26	}	^{+	0.11	}	$	&	$	0.38	_{-	0.02	}	^{+	0.08	}	$	&	$	0.38	_{-	0.08	}	^{+	0.02	}	$											\\[0.2cm]   
$\log \xi_{\rm abs,l}$	&	$	-							$	&	$	1.11	_{-	0.11	}	^{+	0.44	}	$	&	$	0.85	_{-	1.03	}	^{+	0.32	}	$	&	$	<1.3							$	&	$	-3	^f						$											\\[0.2cm]   
CF$_{\rm l}$	&	$	-							$	&	$	> 0.63							$	&	$	0.73	\pm	0.13					$	&	$	0.59	_{-	0.04	}	^{+	0.06	}	$	&	$	0.59	_{-	0.04	}	^{+	0.06	}	$											\\[0.2cm] 
$N_{\rm H,m}$	&	$	-							$	&	$	0.58	\pm	0.09					$	&	$	1.07	^t						$	&	$	0.38	^t						$	&	$	0.38	^t						$											\\[0.2cm]   
$\log \xi_{\rm abs,m}$	&	$	-							$	&	$	1.11	^t						$	&	$	1.5	_{-	0.3	}	^{+	0.2	}	$	&	$	-2.18	^t						$	&	$	-3	^f						$											\\[0.2cm]   
CF$_{\rm m}$	&	$	-							$	&	$	0.74	^t						$	&	$	0.73	^t						$	&	$	0.42	\pm	0.04					$	&	$	0.42	\pm	0.04					$											\\[0.2cm] 
$N_{\rm H,h}$	&	$	-							$	&	$	0.2	_{-	0.05	}	^{+	0.06	}	$	&	$	1.07	^t						$	&	$	0.38	^t						$	&	$	0.38	^t						$											\\[0.2cm]   
$\log \xi_{\rm abs,h}$	&	$	-							$	&	$	1.11	^t						$	&	$	2.15	_{-	0.3	}	^{+	0.2	}	$	&	$	-2.18	^t						$	&	$	-3	^f						$											\\[0.2cm]   
CF$_{\rm h}$	&	$	-							$	&	$	0.74	^t						$	&	$	0.73	^t						$	&	$	0.23	\pm	0.04					$	&	$	0.23	\pm	0.04					$											\\[0.2cm] \hline
$\chi^2/{\rm dof}$	&	$	38.2	/	45					$	&	$	73.4	/	58					$	&	$	68.2	/	58					$	&	$	61.6	/	58					$	&	$	61.6	/	59					$											\\ \hline

\end{tabular}
\begin{tablenotes}
	\item[$f$] fixed.
	\item[$t$] tied.
\end{tablenotes}
\end{threeparttable}
\end{adjustbox}
\label{table:bestfitRelxill}
\end{table}

According to the FFP results, the term $A_{\rm L}(E)x$ in eq.\, (\ref{eq:linear}) can be used to estimate the count rate of the variable component(s) at high energies, for any given count rate, $x$, in the reference band (i.e. the 3--4 keV band). Using this approach,  and the  {\tt FTOOLS} command {\tt ASCII2PHA}, we constructed the spectrum of the variable components when the source flux was lowest, mean, and highest, i.e. when the count rate in the 3-4 keV band was 0.5, equal and 1.5 of the mean (see the dotted, solid and dashed black lines in Fig.\,\ref{fig:lightcurve}). The resulting spectra are plotted in the upper panel of Fig.\,\ref{fig:variableSpec} (circles, triangles and squares represent the low-, mean-, and high-flux spectra, respectively). 

We fitted the variable spectrum above 2 keV with a PL model with a high-energy cutoff ($E_{\rm cut}$), taking into account Galactic absorption only. We assume the same PL slope for the three flux spectra, but we allow different slopes for {\it XMM-Newton} and {\it NuSTAR} data. The best-fit slopes are $\Gamma_{\rm X} = 1.88 \pm 0.02$ and $\Gamma_{\rm N} = 1.94 \pm 0.02$, respectively. We found a 3-sigma lower limit  of 152~keV on $E_{\rm cut}$. The fit is statistically acceptable ($\chi^2/\rm dof=61.9/48,\, p_{\rm null} = 0.09$), however the corresponding residuals reveal an excess in the 12--25\,keV range (see Fig.\,\ref{fig:variableSpec}b). 

This is reminiscent of a reflection component, so we refitted the spectra by the model {\tt TBABS $\times$ RELXILL} in {\tt XSPEC} terminology \citep{wilms00, Daus13,Daus16}. This model includes the PL continuum and a relativistic reflection component. We considered a power-law illumination profile that decreases with distance as $r^{-3}$, we fixed the spin to 0.998 and the reflection fraction to one. We also fixed the inner and outer disc radius to the ISCO and to 400\,$r_{\rm g}$, respectively. All parameters, except normalization, were kept tied among the three flux spectra. The fit is statistically acceptable ($\chi^2/\rm dof = 38.2/45$, $p_{\rm null} = 0.75$), and the best-fit parameters are listed in the second column of Table\,\ref{table:bestfitRelxill}. The corresponding residuals are plotted in Fig.\,\ref{fig:variableSpec}c. Comparing the reflection model to the PL model gives an $F$-test probability of 6.6$\times 10^{-5}$.  Therefore, we find that the observed variations above 2~keV are due to the PL continuum which varies in normalization only, {\it and} to a relativistic reflection component, which varies simultaneously with the continuum (with the reflection fraction remaining constant at the value of one).

\subsection{The variable spectral components at low energies}
\label{subsec:low-variable}

\begin{figure}
\includegraphics[width = \linewidth]{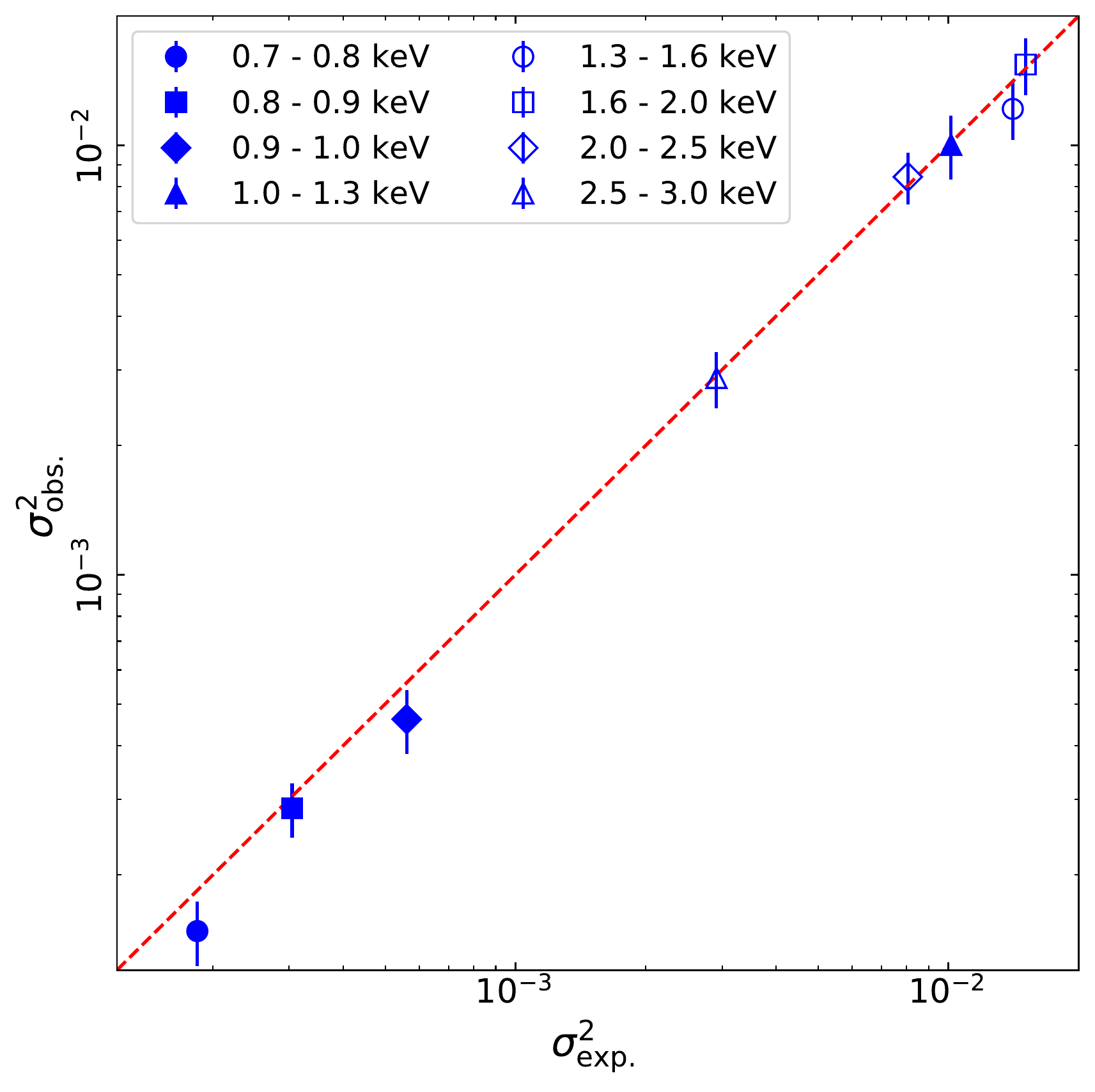}
\caption{Observed versus expected variances in the low-energy bands (see Section \ref{subsec:low-variable}). The red dashed line represents the one-to-one relationship.}
\label{fig:Sigma2}
\end{figure}

It is not certain that the term $V(E)=A_{\rm PLc}(E)x_{3-4 {\rm keV}}^{\beta(E)}$ represents the count rate of the variable component when $E<2$~keV. This is due to the fact that the FFPs at these energies are not well fitted by a straight line\footnote{Although we fitted a PLc model to the FFPs below 3 keV, the best-fit slope of the PLc model at energies between 2 and 3 keV is $\sim 1$, which implies that the FFPs in these energies are in effect well fitted by a straight line}, and it is difficult to interpret the physical meaning of the ``constant" $C_{\rm PLc}$ in the best-fit PLc models. For example, KP17 showed that variable absorption of a PL can result into FFPs with a PLc shape. In this case, the parameter $C_{\rm PLc}(E)$ does not correspond to any physical spectral component (either constant or variable), and the count rate of the variable spectral component is not equal to just $V(E)$ but to $V(E)+C_{\rm PLc}(E)$. In order to investigate whether $V(E)$ represents most of the flux of the variable component(s) at low energies, we performed the following experiment. 

Under the null hypothesis (i.e. $V(E)$ and $C_{\rm PLc}(E)$ are representative of the fluxes of the variable and constant spectral components, respectively), we expect from eq.\,\ref{eq:plc} that
\begin{equation}\label{eq:expecetedSig}
\sigma^2 \left[ y(E)\right] = A_{\rm PLc}(E)\sigma^2\left[x^{\beta(E)} \right],
\end{equation}
\noindent
where $\sigma^2[y(E)]$ is the variance of the light curve in the energy band $E$, and $\sigma^2\left[x^{\beta(E)}\right] = \langle [x^{\beta(E)} - \langle x^{\beta(E)} \rangle  ]^2\rangle$. This is the variance in the 3--4~keV band when instead of $x$, i.e. the count rate in this band, we use $x^{\beta(E)}$. Following the method described in Appendix~\ref{subsec:rms}, we estimated the variance of the light curves in the eight energy bands below 3~keV, $\sigma^2_{\rm obs}[y(E)]$, and the respective expected variances, $\sigma^2_{\rm exp.}[y(E)]$, using eq.\,\ref{eq:expecetedSig} and the best-fit $A_{\rm PLc}(E)$ and $\beta(E)$ values listed in Table\,\ref{table:bestfitPLcXMM}. Fig.\,\ref{fig:Sigma2} shows a plot of the resulting $\sigma^2_{\rm obs}$ versus $\sigma^2_{\rm exp}$ values. Clearly, the observed variations are fully consisted with the variations we would expect if $V(E)$ accounts for most of the flux of the variable spectral component(s) even at energies below 2~keV. 

The points below 2~keV in Fig.\,\ref{fig:variableSpec} show the count rate of the $V(E)$ component at these energies, estimated using the best-fit $A_{\rm PLc}$ and $\beta$ values listed in Table\,\ref{table:bestfitPLcXMM}. The extrapolation of the best-fit model presented in the previous section to energies below 2~keV reveals clear residuals, characterized by a dip in the 1-1.3\,keV band whose strength varies with flux: the residuals get smaller at higher (intrinsic) flux (see Fig.\,\ref{fig:variableSpec}c). These are indicative of a variable absorption affecting the soft X-rays. We therefore fitted the 0.7-40\,keV spectra by the model {\tt TBABS $\times$ ZXIPCF $\times$ RELXILL}, where {\tt ZXIPCF} \citep[][]{zxipcf} is a partial covering absorption model. We tried three different configurations to account for the absorption variability. First we let the column density ($N_{\rm H}$) free to vary among the three spectra, and keep the ionization parameter ($\log \xi_{\rm abs}$) and the covering fraction (CF) tied (hereafter model A). Then we let $\log \xi_{\rm abs}$ free to vary and kept $N_{\rm H}$ and CF tied (hereafter model B). Finally, we let CF variable and kept $N_{\rm H}$ and $\log \xi_{\rm abs}$ tied (hereafter model C). The resulting best-fit parameters are listed in Table\,\ref{table:bestfitRelxill}. Given the low ionization of the absorber in model C, we re-fitted the spectra by fixing $\log \xi_{\rm abs}$ to -3 (hereafter model C$'$). The fit is statistically comparable to model C ($\chi^2/\rm dof = 61.6/59$) and the best-fit parameters (listed in Table\,\ref{table:bestfitRelxill}) are consistent between the two models. The fit was statistically acceptable in all three cases  ($p_{\rm null} = 0.08, 0.17$ and 0.35, respectively). {The difference between the best-fit $\Gamma_{\rm X}$ and $\Gamma_{\rm N}$ is $\Delta \Gamma = 0.1 \pm 0.03$. \cite{Madsen2015} studied the the cross-calibration uncertainty between EPIC-pn and FPMA/B for PKS 2155-304 and 3C 273. The authors reported a difference in slopes $\Delta \Gamma = 0.02 \pm 0.09$ and $0.04 \pm 0.04$, respectively. The weighted mean of the differences that they found, $\overline{\Delta \Gamma} = 0.04 \pm 0.04$, is statistically consistent with the value we obtained from model C$'$ within $\sim 1.2~\sigma$. Despite the fact that the spectral slopes of EPIC-pn and {\it NuSTAR} are statistically consistent with each others, our results suggest that the slopes inferred from the {\it NuSTAR} spectra are softer than the ones obtained from the EPIC-pn. We note that similar differences between EPIC-pn and FPMA/B slopes have been recently reported in the literature \citep[e.g.][]{Cappi16, Ballo17}.}
 
In order to test which model describes best the variable spectrum we compared models B and C$'$ (they provide the two lowest $\chi^2$) we computed the Akaike information criterion, ${\rm AIC_c}$, corrected for the bias introduced by the finite size of the sample \citep{Akaike73, Sugiura78},
\begin{equation}\label{eq:aic}
{\rm AIC_c} = 2k - 2C_{\rm L} + \chi^2 + \frac{2k(k+1)}{N-k-1}.
\end{equation}
\noindent
$C_{\rm L}$ is the constant likelihood of the true hypothetical model (it does not depend neither on the data nor on the tested models), $k$ is the number of free model parameters, and $N$ is the number of data points in the spectra. The model with the lowest ${\rm AIC_c}$ should be the most preferred one among the two models. We then computed the differences $\Delta [{\rm AIC_c}]_{\rm B} = {\rm AIC_{c,\, B}} - {\rm AIC_{c,\,C'}}$ and $\Delta [{\rm AIC_c}]_{\rm C'} = {\rm AIC_{c,\, C'}} - {\rm AIC_{c,\,B}}$, and the `Akaike weight', 
\begin{equation}
\rm W[AIC_c]_B = \frac{e^{-0.5\Delta [AIC_c]_{\rm B}}}{e^{-0.5\Delta [AIC_c]_{C'}} + e^{-0.5\Delta [AIC_c]_B}}.
\end{equation}
\noindent
This weight provides a measure of the `strength of evidence' for model B. We found a low value $\rm W[AIC_c]_B = 0.7\%$, which means that the variable ionization model has less than 1 per cent chance of being the best model among the two scenarios. Hence we consider model C$'$ (i.e. the case of a neutral absorber, with constant $N_{\rm H}$ and variable CF) as the most probable model to explain the observed variations at low energies.

\subsection{The constant component}
\label{subsec:highE-constant}
\begin{table}
\centering
\caption{Best-fit parameters obtained by fitting the spectrum of the constant components.}
\begin{threeparttable}
\begin{tabular}{lcc}
\hline  
\multicolumn{3}{c}{\tt ZBBODY}													\\[0.1cm]   
$kT_{\rm BB}$ (keV)	&	$	0.2	\pm	0.02	$	&	$	0.18	\pm	0.01	$	\\[0.1cm]   
Norm$ \times 10^{-4}$	&	$	1.7	\pm	0.2	$	&	$	4.4	\pm	0.6	$	\\[0.1cm]   
\multicolumn{3}{c}{\tt ZXIPCF}													\\[0.1cm]   
$N_{\rm H} \times 10^{22}\,\rm cm^{-2}$	&	$	-			$	&	$	0.38^f			$	\\[0.1cm]   
$\log \xi$	&	$	-			$	&	$	-3^f			$	\\[0.1cm]   
CF	&	$	-			$	&	$	0.42^f			$	\\[0.1cm]     
\multicolumn{3}{c}{\tt PEXMON}													\\[0.1cm]   
$\Gamma$	&	$	2.16	\pm 0.2	$	&	$	2.16	_{-0.1}	^{+0.2}	$	\\[0.1cm]   
$E_{\rm cut}^\dagger$\,(keV)	&	$	>20			$	&	$		>20		$	\\[0.1cm]   
$A_{\rm Fe}$\,solar	&	$	1.2	\pm	0.3	$	&	$	1.2	\pm	0.3	$	\\[0.1cm]   
$i(^\circ)$	&		30$^f$				&	$	30^f			$	\\[0.1cm]   
Norm $\times 10^{-3}$	&	$	7.6	\pm	3.5	$	&	$	7.6	\pm	3.5	$	\\[0.1cm]   \hline
$\chi^2 / {\rm dof}$	&	$		13/17		$	&		$13/17$				\\[0.1cm]   \hline
\end{tabular}
\begin{tablenotes}
	\item[$\dagger$] three-sigma lower limit.
	\item[$^f$] fixed.
\end{tablenotes}
\end{threeparttable} 
\label{table:bestfitpexmon}
\end{table}
\begin{figure}
\centering
\includegraphics[width= 0.95\linewidth]{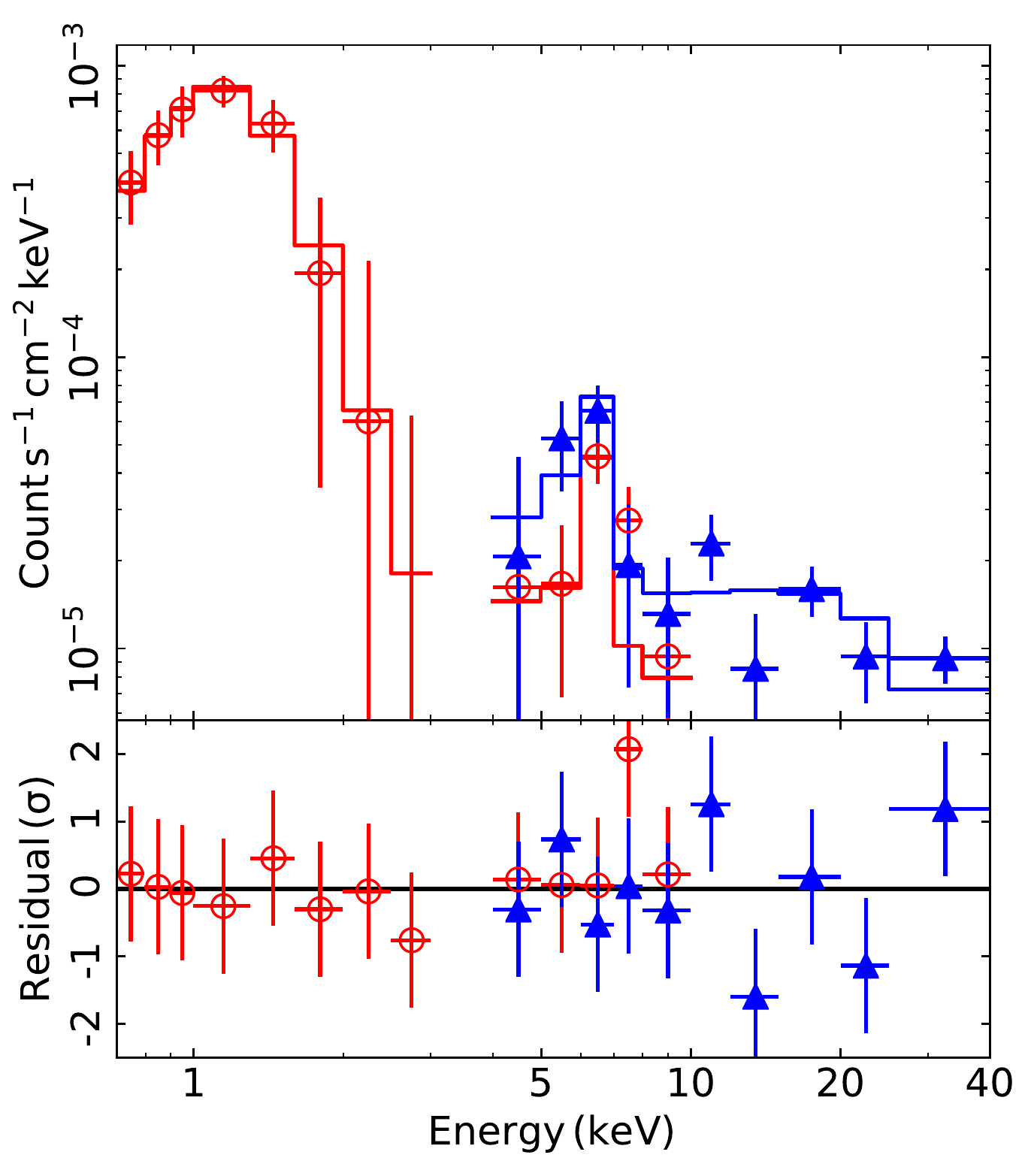}
\caption{The full-band spectrum of the constant flux components, fitted with a BB plus neutral reflection model taken into account Galactic absorption only.}
\label{fig:constantspec}
\end{figure}

The constant terms in the linear and PLc models for the high and low energy bands, respectively, should be representative of the count rate of non-variable spectral components. As above, we used the best-fit $C_{\rm L}$  and $C_{\rm PLc}$ values listed in Tables \ref{table:bestfitlinearXMM}-\ref{table:bestfitPLcXMM} and the {\tt FTOOLS} command {\tt ASCII2PHA}, and we constructed the spectrum of the constant-flux (hereafter, simply constant) spectral components. the resulting spectrum is plotted in the upper panel of Fig.\,\ref{fig:constantspec}. 

We fitted the resulting spectrum with the sum of a blackbody (BB) and neutral reflection \citep[{\tt {PEXMON}};][]{Nandra07} components, taking into consideration Galactic absorption only (model {\tt TBABS(ZBBODY + PEXMON)} in XSPEC terminology). We let both the BB temperature ($kT_{\rm BB}$) and the normalization free to vary. We fixed the reflection fraction of the {\tt PEXMON} component to {minus one (so that we account only the reflection component, assuming an isotropic emission from the primary source)}, and the abundance of heavy elements to solar but we let the iron abundance, inclination, PL slope and $E_{\rm cut}$ energy free to vary. {We remind the reader that the values of the constant from {\it NuSTAR} are estimated by adding the light curves from FPMA and FPMB. For that reason we multiplied the aforementioned model by a constant fixed to 1 for {\it XMM-Newton} and 2 for {\it NuSTAR}.} The model fits well the data, however we could not get constrains on the inclination of the reprocessing material, thus we fixed it to the best-fit value obtained in model C$'$ $i = 30^\circ$ and refitted the model. The best-fit is statistically accepted ($\chi^2/\rm dof = 13/17$). The best-fit parameters are listed in the second column of Table\,\ref{table:bestfitpexmon}, and the best-fit model is shown in Fig.\,\ref{fig:constantspec}.

We have also checked the possibility that the variable, intrinsic absorber affects the BB. We therefore re-fitted the data with the model: {\tt TBABS(ZXIPCF $\times$ ZBBODY + PEXMON)}. We fixed the column density, ionization and covering fraction of the absorber to 3.8$\times 10^{21}\rm cm^{-2}$, -3 and 0.42, respectively (these are the best-fit values obtained for the the mean-flux spectrum of the variable components in the previous section). The fit is statistically good ($\chi^2/\rm dof = 13/17$). The best-fit parameters, listed in the last column of Table\,\ref{table:bestfitpexmon}, are all in agreement with the ones obtained from an unabsorbed BB, except for the normalization of the BB component that was found to be higher than the unabsorbed case.

\section{Energy Spectra}
\label{sec:energy_spectra}

\begin{figure}
\centering
\includegraphics[width=\linewidth]{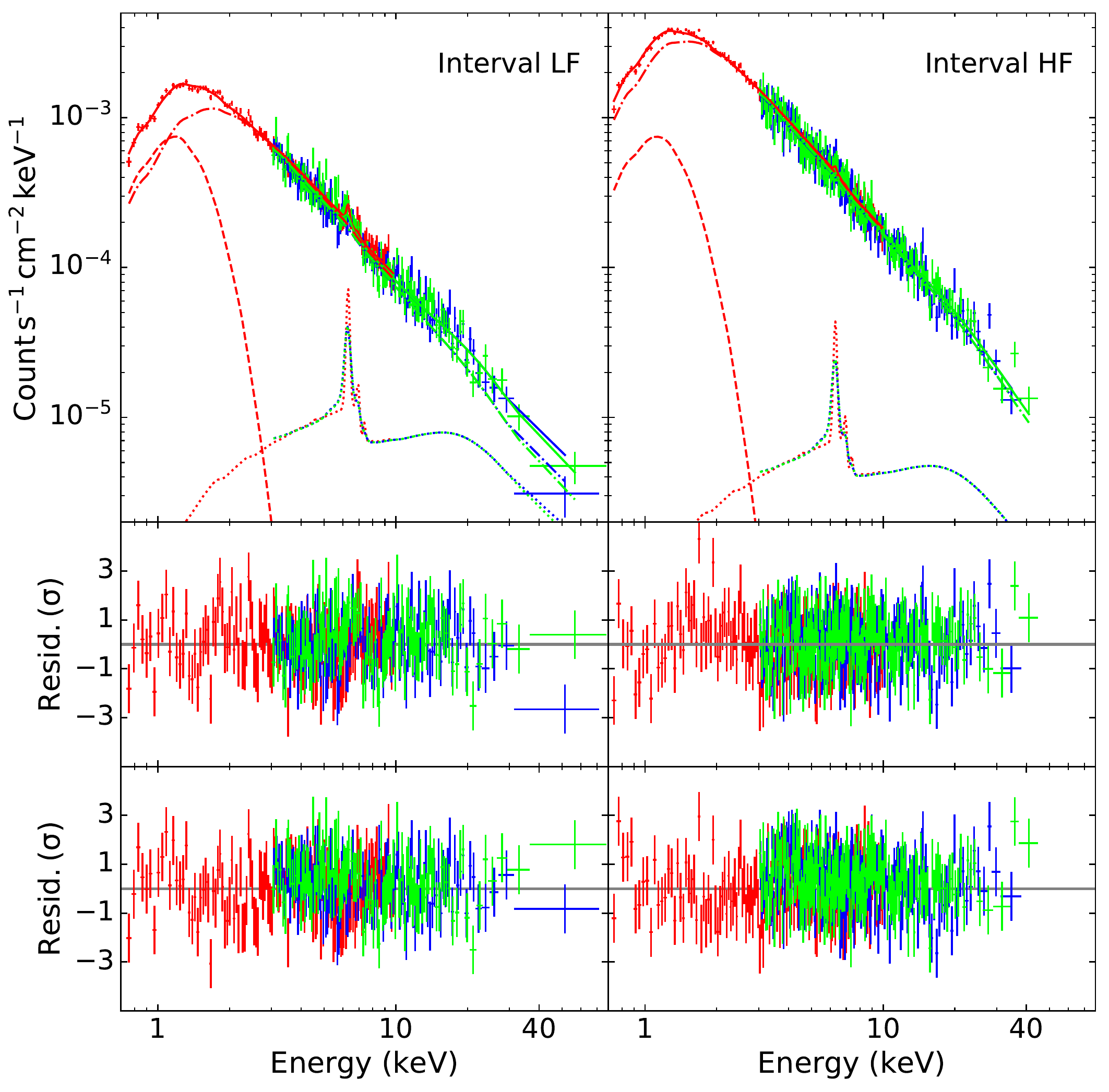}

\caption{Top panels: The EPIC-pn (red) and FPMA,B (blue, green) LF and HF spectra (left and right panel, respectively). The dashed-dotted, dash and dotted lines correspond to the best-fit PL+ionised reflection, BB and neutral reflection components, respectively (see Section \ref{sec:energy_spectra} for details). Middle panels: the best-fit residuals. Bottom panels: the residuals obtained by fitting the corresponding spectra with $\rm model_{Ref.}$ (see Section \ref{sec:conclusion} for details).}
\label{fig:SpectralFit}
\end{figure}

To test the results obtained from the FFP analysis,  we extracted the source spectra from the 12-ks-long intervals shown in Fig.\,\ref{fig:lightcurve}. These two intervals correspond to the source emitting the lowest and highest flux, and we will denote them as the LF and HF intervals, respectively. The corresponding spectra are shown in the top panel of Fig.\,\ref{fig:SpectralFit}. 

According to our FFP analysis results (Section \ref{sec:results}) the source spectrum should be well fitted by an intrinsically absorbed PL plus ionised reflection, neutral reflection and a BB component, i.e. by a model of the form ${\tt TBABS \left[ ZXIPCF \times (RELXILL + ZBBODY) + PEXMON \right]}$ in XSPEC terminology. We fixed all the parameters, except the normalisations, to their best-fit values reported in Tables\,\ref{table:bestfitRelxill} and \ref{table:bestfitpexmon}. The covering fraction of the absorber (assumed to be neutral) were kept frozen at 0.59 and 0.23 for the LF and HF spectra, respectively, and we fitted them separately. The model fitted both spectra very well ($\chi^2/\rm dof = 390/373$ and $528/518$, for the LF and HF spectra, respectively). The best-fit normalisation of the BB and of the neutral reflection components are consistent with the ones reported in Table\,\ref{table:bestfitpexmon}. The best-fit model residuals are shown in the middle panel of Fig.\,\ref{fig:SpectralFit}. Interestingly, a similar model where the BB component is not affected by the intrinsic absorber does not fit well the LF spectrum ($\chi^2/\rm dof = 446/373$, $p_{\rm null} = 0.005$). We conclude that the FFP results regarding the variable and constant spectral components in the X--ray spectrum of the source are consistent with the observed energy spectrum. 

\section{Discussion and conclusions}
\label{sec:conclusion}

We studied the spectral variability of SWIFT\,J2127.4+5654 using three simultaneous {\it XMM-Newton} and {\it NuSTAR} observations. The observations allowed us to study the spectral variations on time scales from 1-5.8~ks up to $\sim 5.2$~d, and over a broad energy range, from $\sim 0.7$ up to $\sim 40$ keV. By studying the FFPs, we were able to identify the variable and the constant spectral components in the source, in a model-independent way. The results from the FFP analysis are in full agreement with the observed energy spectra of the source. We summarize below our main findings.

a) The variations at energies above 2~keV are due to a PL primary component, which varies in normalisation only. Absorption related variations are ruled out in these energies. We find strong evidence for a reflection component, which varies simultaneously with the PL component. We cannot constrain its  parameters, but they are consistent with a relativistic reflection from the inner disc, which is illuminated by an X-ray source on top of it. The fact that the reflection component must vary simultaneously with the continuum, reinforces this hypothesis. 

b) There is an additional variable component at energies below 2~keV. We identify this component with a neutral absorber of $N_{\rm H} = 3.8\times 10^{21}\,\rm cm^{-2}$ whose covering fraction varies between $\sim 0.2$ and $\sim 0.6$. 

c) We also detect spectral components whose flux does not vary over the entire length of the observations. They are consistent with reflection from neutral material, and with a BB of temperature $\sim 0.18$~keV. 

\noindent We discuss the implications of our results below. 

\subsubsection*{The variable components at energies $>2$~keV}

The fact that the FFPs in the $\sim 2-40$~keV range are well fitted with a straight line is a simple yet powerful and constraining result. It rules out the possibility of either spectral slope and/or absorption variations contributing significantly to the variability of the source at these energies. Any of these variability scenarios would then result in non-linear FFPs \citep[see][KP17]{Kammoun15}.

We estimated the spectra of the variable component at the lowest, mean and highest flux levels of the source. A simple PL model is able to fit well these spectra but the residuals suggest the presence of a second component. A model consisting of a variable PL plus reflection component improved the fit significantly. The key issue here is the fact that the reflection component must vary simultaneously with the primary emission, on timescales as short as 1~ks. This is in agreement with the high-frequency lags reported by \citet{Marinucci14Swift} and \cite{Kara15}. Such a quick responding component implies reflection from material located close to the central source, and supports the idea of relativistic reflection from the inner disc.  

\subsubsection*{The variable components at energies $<2$~keV}

We find strong evidence for the presence of an absorber which is intrinsic to the source, varies on short time scales and affects its spectrum at energies below $\sim 2$ keV. We investigated the possibility of the absorber varying in a) column density, b) ionisation, or c) covering fraction. All models fit well the spectra of the variable components (plotted in Fig.\,\ref{fig:variableSpec}), but the model where the covering fraction anticorrelates with the source flux is the most favoured scenario, from a statistical point of view. 

However, we cannot provide an explanation for this variability scenario, i.e. why the covering fraction of this absorber would correlate with the (intrinsically) variable flux of the continuum. If the absorber is associated with some kind of wind from the disc, perhaps the intrinsic, continuum flux variations may be associated with the amount of material that is expelled from the disc, but a detailed investigation of such assumptions are beyond the scope of this work. A more physically plausible scenario would be a variability in the ionisation state of the absorber which could, naturally, correlate with the source flux. This scenario fits well the variable component spectra, however it is statistically less favoured, based on the AIC, when compared to the model with a variable covering fraction. 

We note that the full band (i.e. 0.7-- 40 keV) spectra of the variable components (Fig\,\ref{fig:variableSpec}) cannot be fitted by a ``PL plus a variable absorber" model. We find a $\chi^2$ of 91.7, for 62 degrees of freedom in this case. The quality of the fit is rather poor ($p_{\rm null}=8.5\times 10^{-3}$), and is certainly much worse than the quality of the model C$'$ fit to the same data set (best-fit results listed in the last column of Table \,\ref{table:bestfitRelxill}). A simple $F-$test indicates a significant improvement on the best-fit quality ($p_{\rm null}=3\times 10^{-5}$ when we add the variable reflection component. 

\subsubsection*{The constant components}

We find evidence for the presence of spectral components which remain constant over $\sim 5$~days (at least). The spectrum of these components can be well described by a BB component ($kT_{\rm BB} \simeq 0.18$~keV) dominating at energies below $\sim 2$~keV, and a neutral reflection dominating at higher energies. The fact that this reflection component is constant suggests that the material responsible should be located far from the central source. These components can account for $\sim 10-25$\% of the total observed flux of the source, in the 0.7--40~keV range.

The origin of the BB-like emission is not clear. A hint is provided by the fact that this component may be affected by the variable (intrinsic) absorber, as the results from the fits to the low- and high-flux spectra of the source showed (see Section \ref{sec:energy_spectra}). In this case the material emitting this component should be located close to the X--ray source, hence, this component may be the result of intrinsic emission from the inner disc. Similarly to KP17, we tested this idea by fitting the spectrum of the constant components (Fig.\,\ref{fig:constantspec}) with the {\tt OPTXAGNF} model \citep{Done12}. We considered the emission from an accretion disc around a maximally rotating Kerr BH with a mass of $1.5\times 10 ^7 \,\rm M_\odot$, and we fitted the model to the spectrum below 3~keV. The model could not fit the data if we consider the Galactic absorption only. Instead, by adding an intrinsic neutral absorption, with a column density fixed at $N_{\rm H} = 3.8\times 10^{21}\,\rm cm^{-2}$ (as reported in Table\,\ref{table:bestfitRelxill}), the fit is statistically accepted ($\rm \chi^2/dof = 12.1/6$, $p_{\rm null}= 0.06$). The best-fit suggests a BH accrreting at a super-Eddington luminosity ($\log (L/L_{\rm Edd}) = 1.01 \pm 0.02$). 

As for the neutral reflection, being constant for a $\sim 5.2$~d indicates that the reflecting material is located at a distance larger than $\sim 1.35\times10^{16}$~cm. This is in agreement with the results by \citep{Sanfrutos13} who identified the presence of neutral material, associated with the broad-line region, at a distance larger than $\sim 4.3\times 10^{16}$~cm in this source.

\subsubsection*{FFPs versus energy spectra}

It is well known that different models can fit well the time average spectra of AGN, and it is becoming more obvious in the last few years that it is only through the combined spectral and the timing study of the data we can break the degeneracy between various models and identify, correctly, the variable spectral components in these objects. 

To demonstrate this issue, we re-fitted the LF and HF spectra (see Section $\ref{sec:energy_spectra}$) with a model similar to the one considered by \cite{Marinucci14Swift}. These authors fitted the time-average spectrum of each one of the three observations of the source, using the same {\it XMM-Newton} and {\it NuStar} data that we used. They found that these spectra could be fitted well by a model consisting of an intrinsically absorbed PL plus blurred reflection component which vary in flux only, plus a neutral, constant reflection component. They found the intrinsic absorber to be constant, and non-ionized, with a column density of $N_{\rm H} = 2.13\times 10^{21}~\rm cm^{-2}$.  

The \cite{Marinucci14Swift} model in {\tt XSPEC} terminology can be written as $ {\tt TBABS (ZTBABS \times RELXILL + XILLVER),}$ where {\tt TBABS} and {\tt ZTBABS} correspond to the Galactic and intrinsic absorption, respectively. Like \cite{Marinucci14Swift}, we kept the ionisation in {\tt XILLVER} fixed to its lowest allowed value ($\log \xi = 0$), we tied the photon index, $E_{\rm cut}$ and inclination of {\tt RELXILL} and {\tt XILLVER}, we fixed the emissivity index of {\tt RELXILL} to 6, and we considered the accretion disc extending from the ISCO up to 400~$\rm r_g$. This model fits well the LF and HF spectra ($\chi^2/\rm dof = 380/367$ and $530/513$, respectively). We note that, in both cases, the intrinsic absorber is not needed, statistically, by the fit, and the best-fit parameters are consistent with the values reported by \cite{Marinucci14Swift}. We show the best-fit residuals in the bottom panel of Fig.\,\ref{fig:SpectralFit}.
 
So, both the \cite{Marinucci14Swift} and the model we described in Section \ref{sec:energy_spectra} can fit the LF and HF spectra equally well. However, the \cite{Marinucci14Swift} model cannot explain many of the FFP analysis results. For example, the fact that the low-energy FFPs are significantly higher than the FFPs which we would expect in the case of a PL varying in normalization only (compare the data with the dashed lines in Fig.\ref{figapp:lowEFFPs}) clearly shows, without any additional modelling, that there exists an extra constant and/or variable component at low energies, {\it in addition} to the absorbed, variable PL+reflection component. {According to our analysis the extra flux at low energies is due to the non-variable black body component that we studied in Section \ref{subsec:highE-constant}.}

There are many ways we can perform a timing study of the AGN, X--ray data, the simplest of which is the study of the ``root mean square" (RMS) spectra \citep[see e.g.][]{Edelson2002}, and the study of the FFPs. Both methods provide the same information. The FFP analysis is a relatively easy method to identify the various spectral components in the X--ray spectra of AGN, unambiguously. It can be used to study the variable as well as the constant spectral components of the source and we believe that, a combined study of the energy spectrum and of the FFPs can provide give a relatively complete picture of the X--ray spectral variability in the case of X--ray bright, and variable AGN. 

\section*{Acknowledgements}

ESK acknowledges the support by the ERASMUS+ Programme - Student Mobility for Traineeship - Project KTEU - ET (Key to Europe Erasmus Traineeship). ESK thanks the Department of Physics at the University of Crete for warm hospitality, where this project was finalised. This work made use of data from the {\it NuSTAR} mission, a project led by the California Institute of Technology, managed by the Jet Propulsion Laboratory, and funded by NASA, {\it XMM-Newton}, an ESA science mission with instruments and contributions directly funded by ESA Member States and NASA. This research has made use of the {\it NuSTAR} Data Analysis Software (NuSTARDAS) jointly developed by the ASI Science  Data Center (ASDC, Italy) and the California Institute of Technology (USA). The figures were generated using {\tt matplotlib} \citep{Hunter07}, a {\tt PYTHON} library for publication of quality graphics.

\bibliographystyle{mnras}
\bibliography{ek-Swift-ref} 

\appendix

\section{Plots}
\label{app:plots}

\begin{figure}
\centering
{\includegraphics[width=0.235\textwidth]{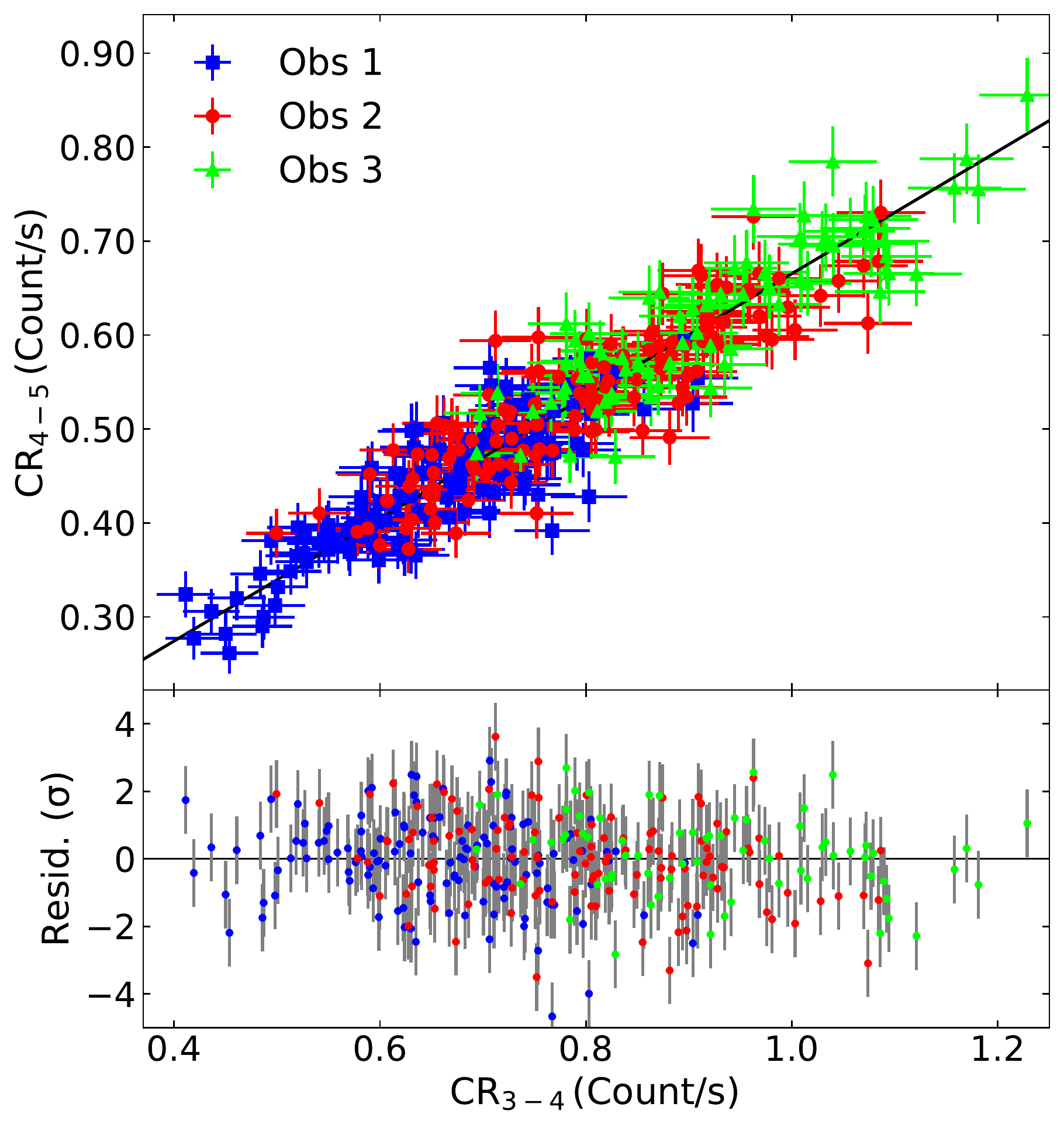}}
{\includegraphics[width=0.235\textwidth]{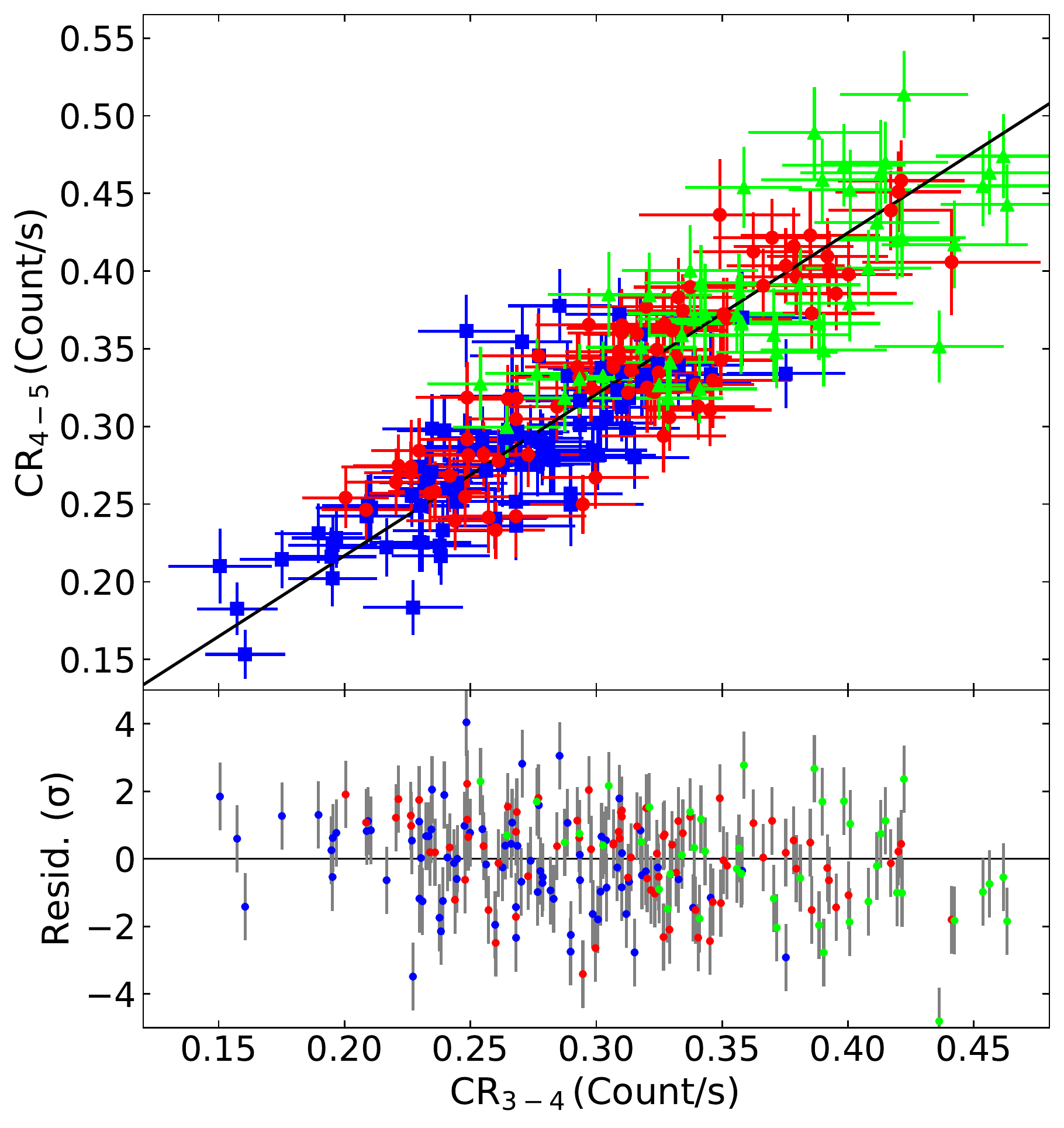}}
{\includegraphics[width=0.235\textwidth]{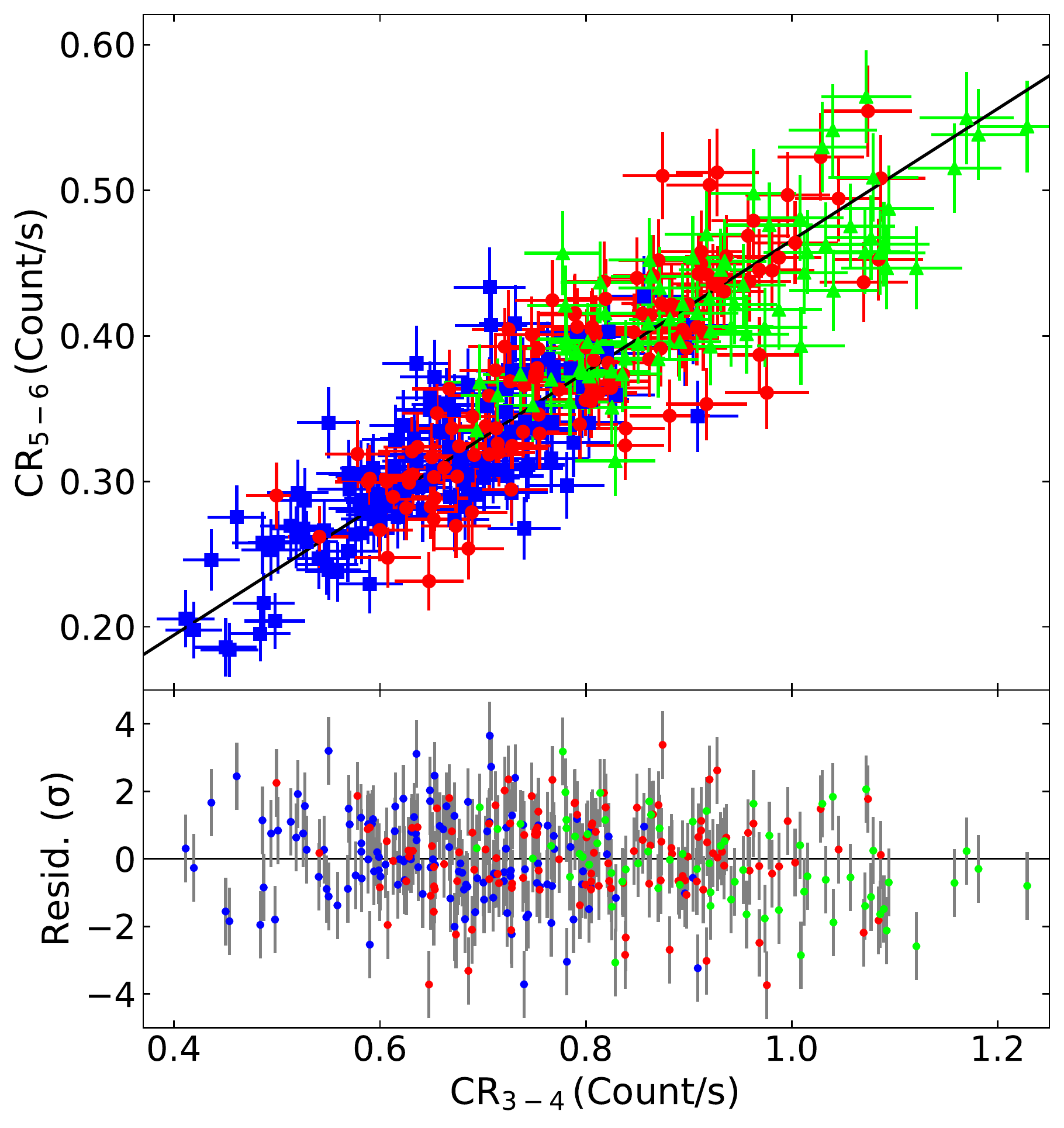}}
{\includegraphics[width=0.235\textwidth]{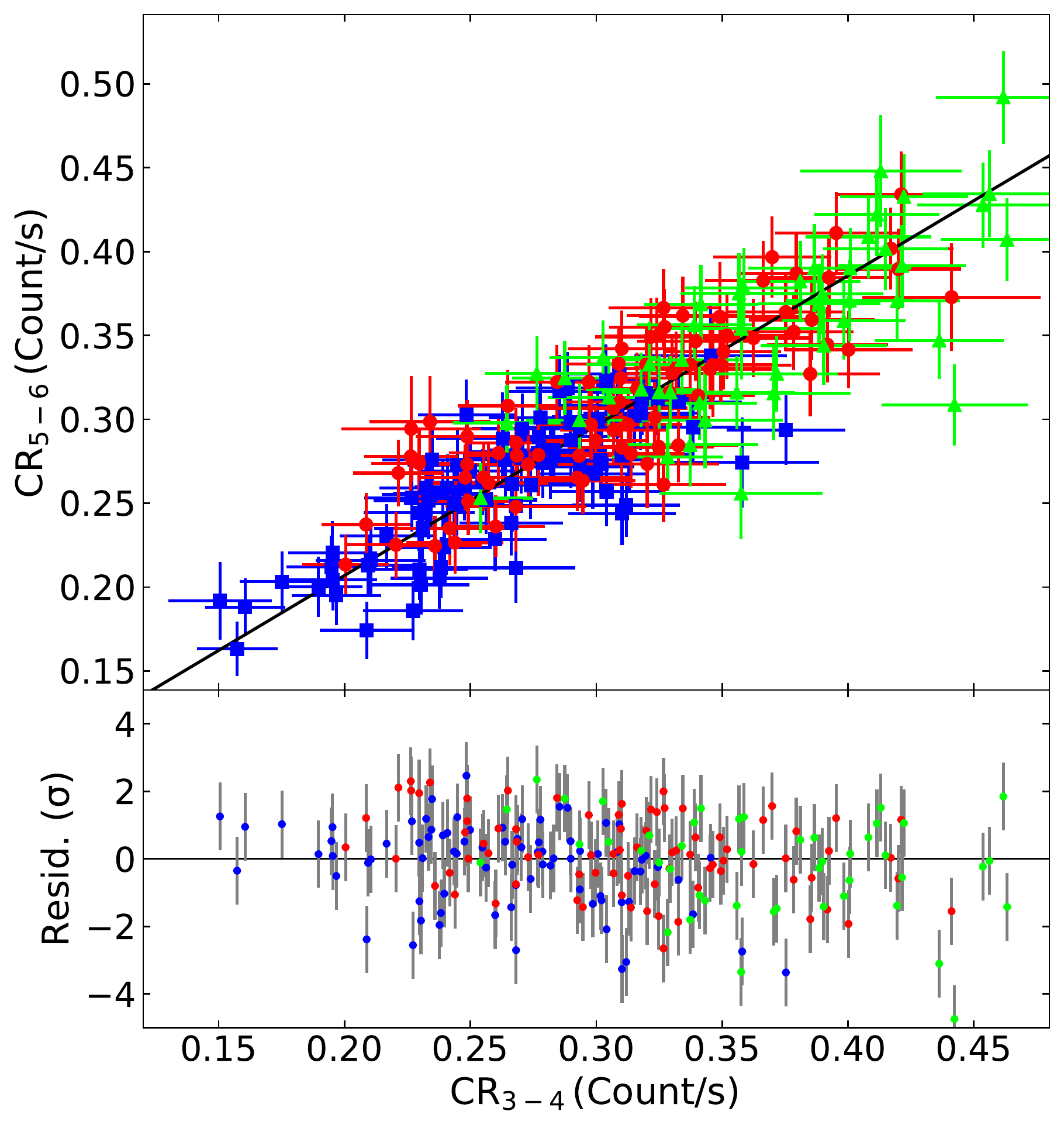}}
{\includegraphics[width=0.235\textwidth]{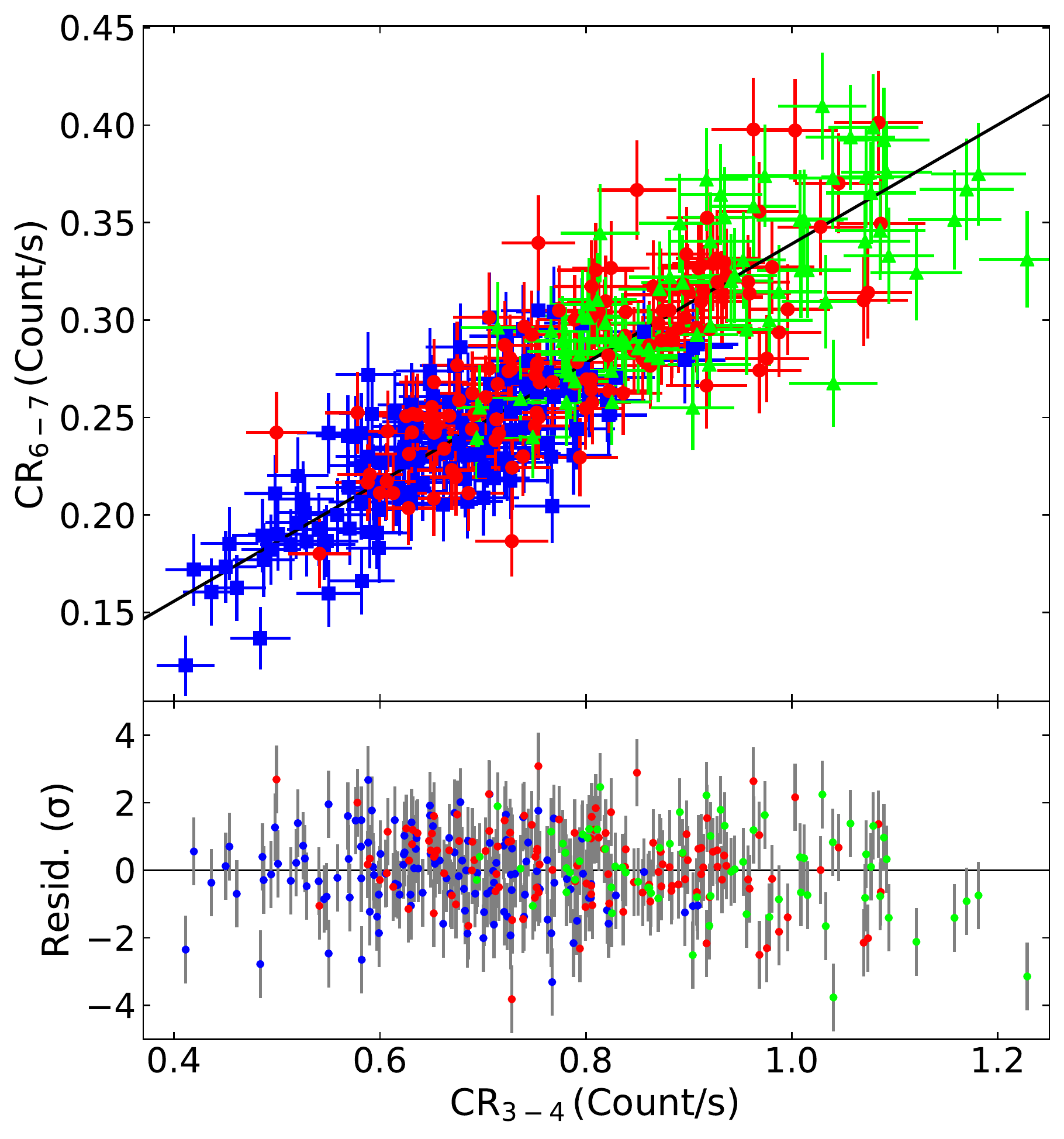}}
{\includegraphics[width=0.235\textwidth]{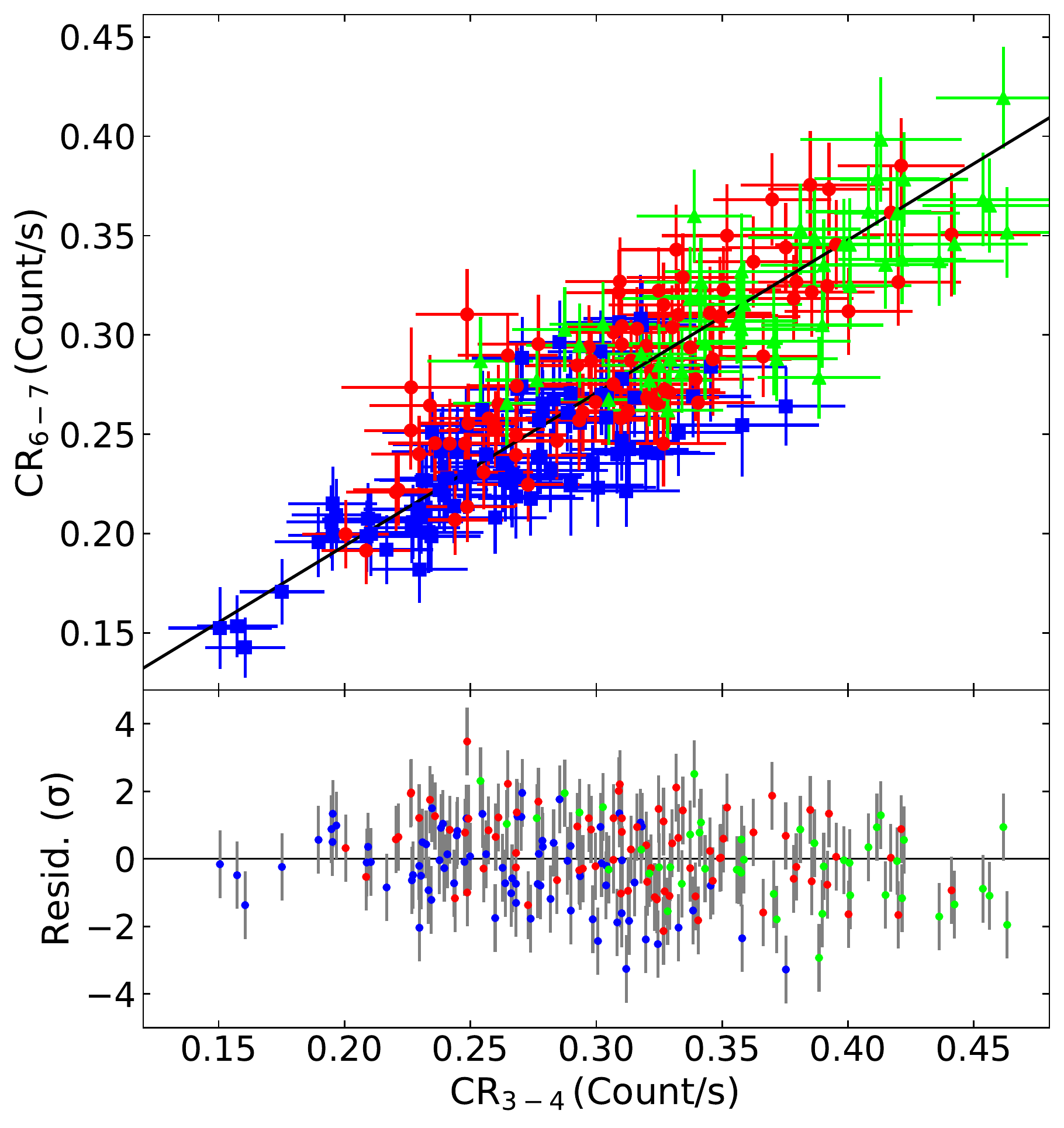}}
{\includegraphics[width=0.235\textwidth]{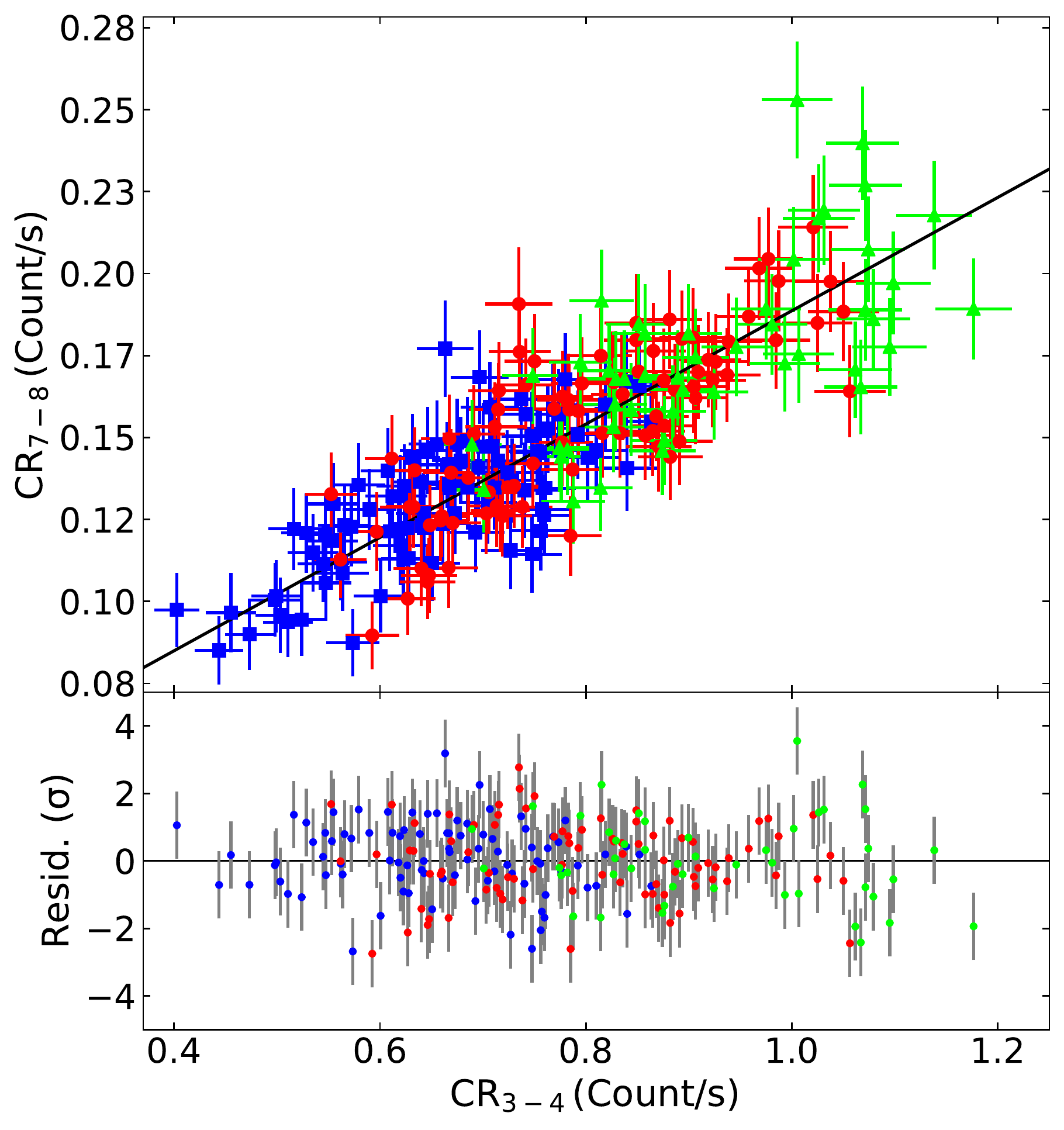}}
{\includegraphics[width=0.235\textwidth]{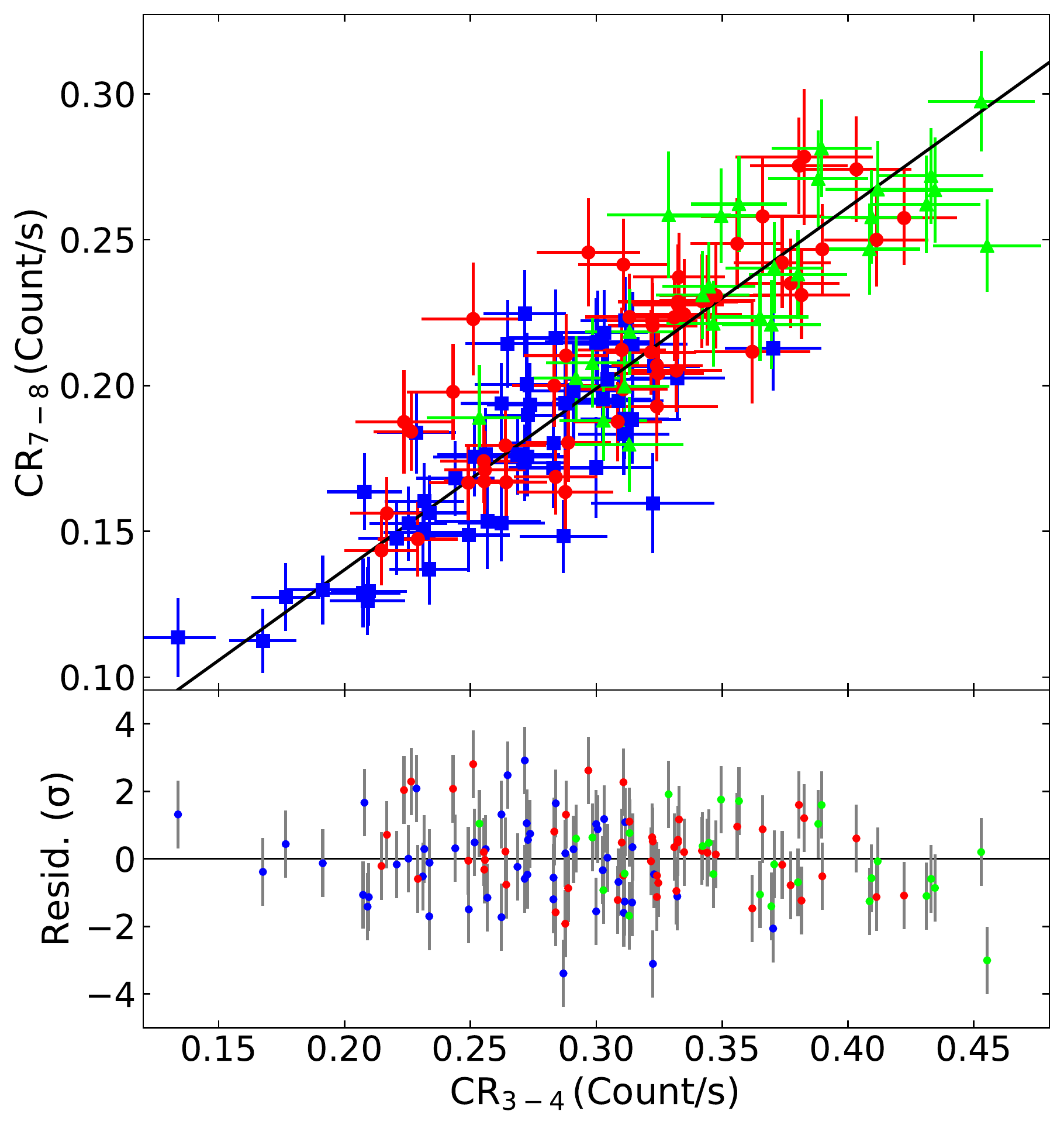}}
{\includegraphics[width=0.235\textwidth]{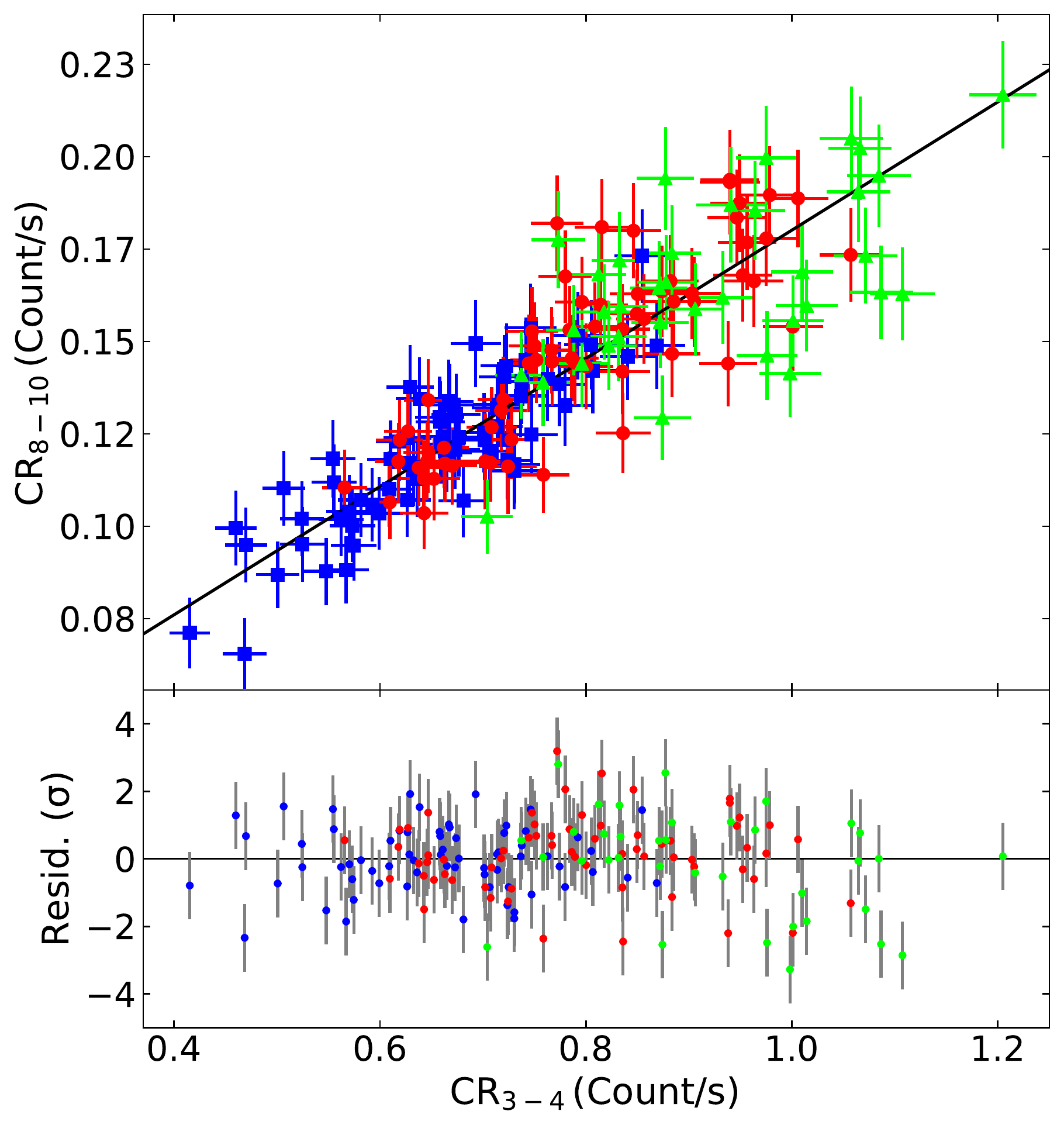}}
{\includegraphics[width=0.235\textwidth]{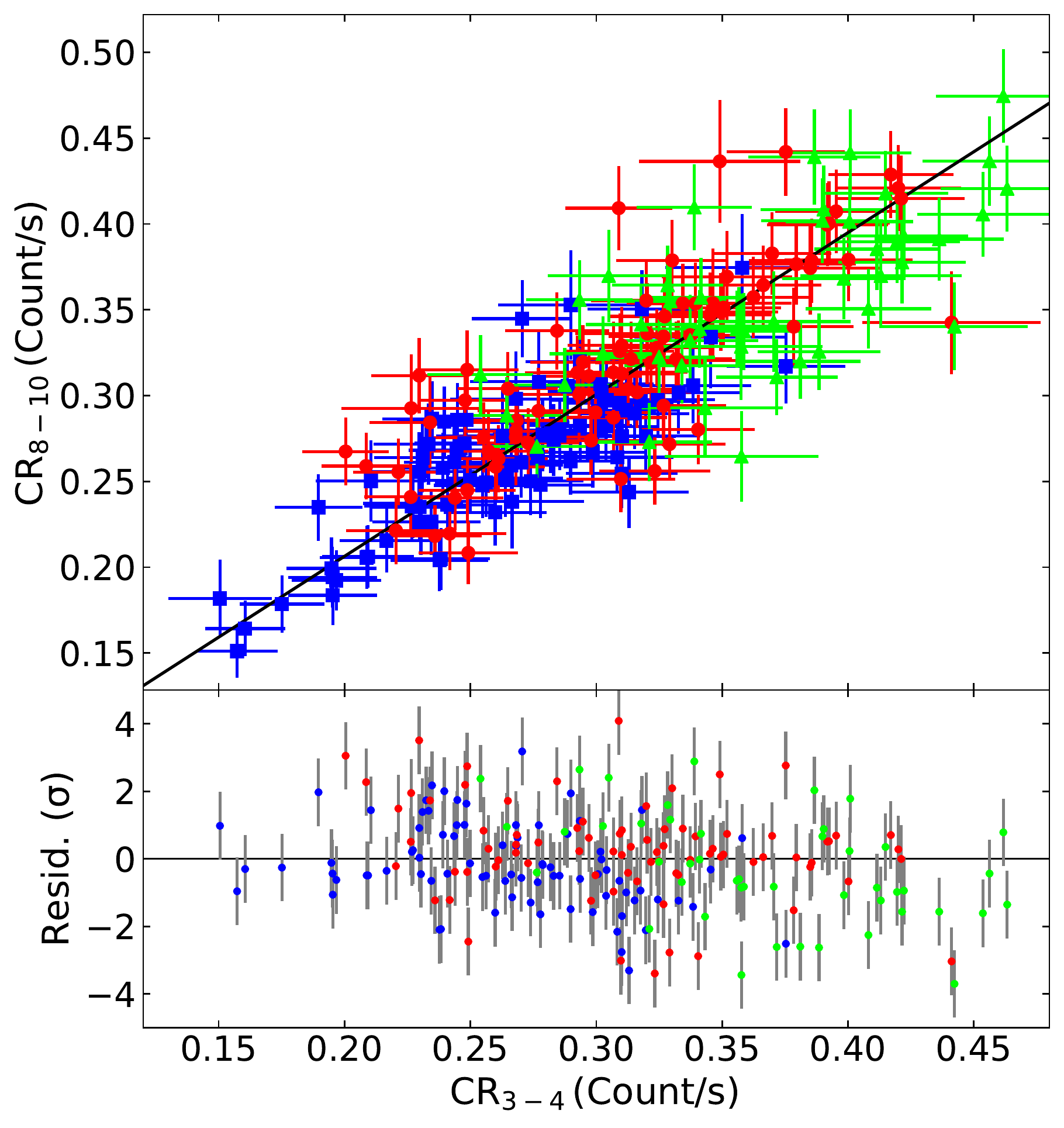}}
\caption{{\it XMM-Newton} and {\it NuSTAR} (left and right column, respectively), high-energy FFPs in the common energy bands (4--10\,keV). The solid black line indicates the best-fit linear model to the combined FFPs. Best-fit residuals are plotted in the lower panel of each plot. }
\label{figapp:commFFPs}
\end{figure}

\newpage

\begin{figure}
\centering

{\includegraphics[width=0.235\textwidth]{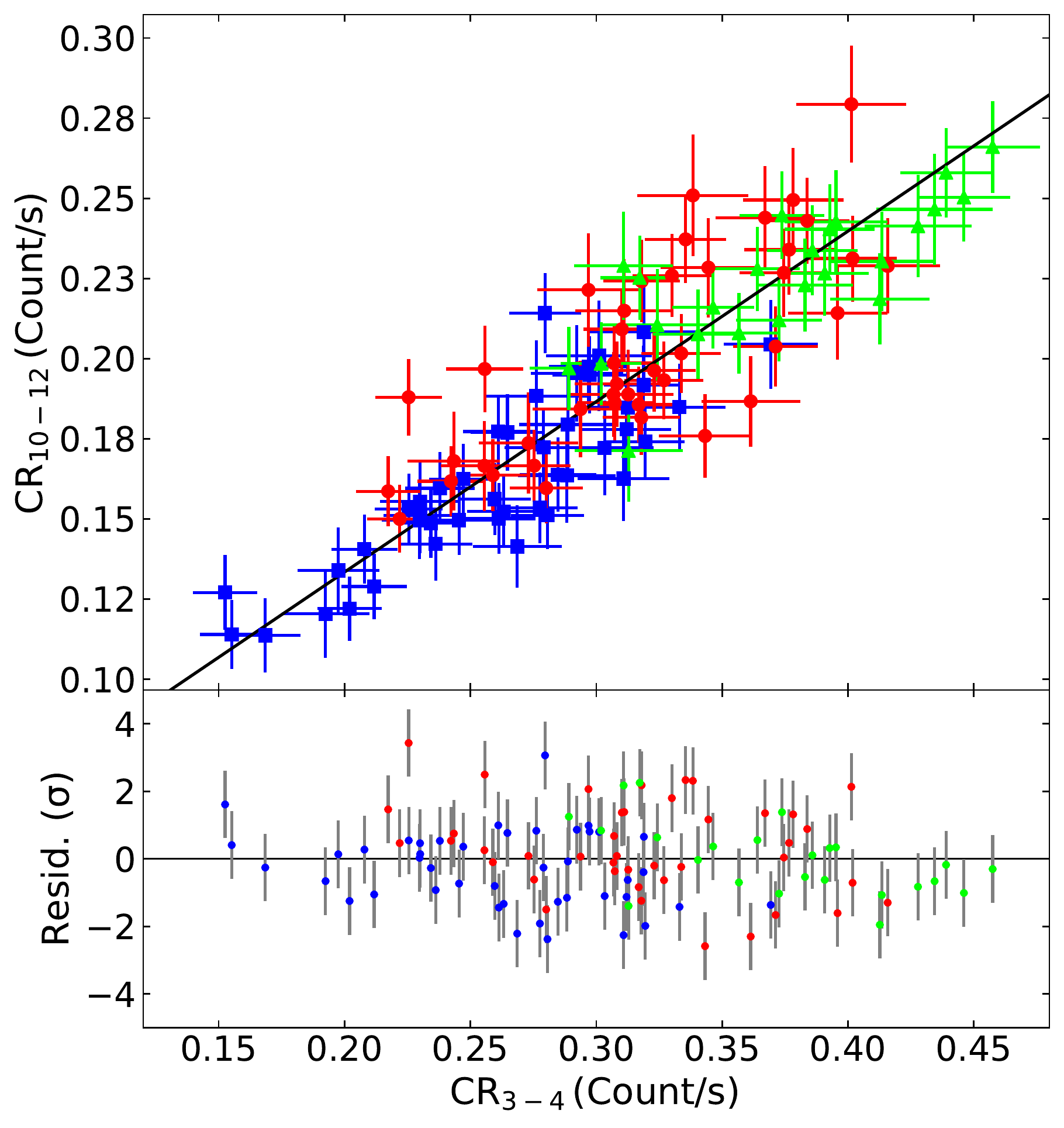}}
{\includegraphics[width=0.235\textwidth]{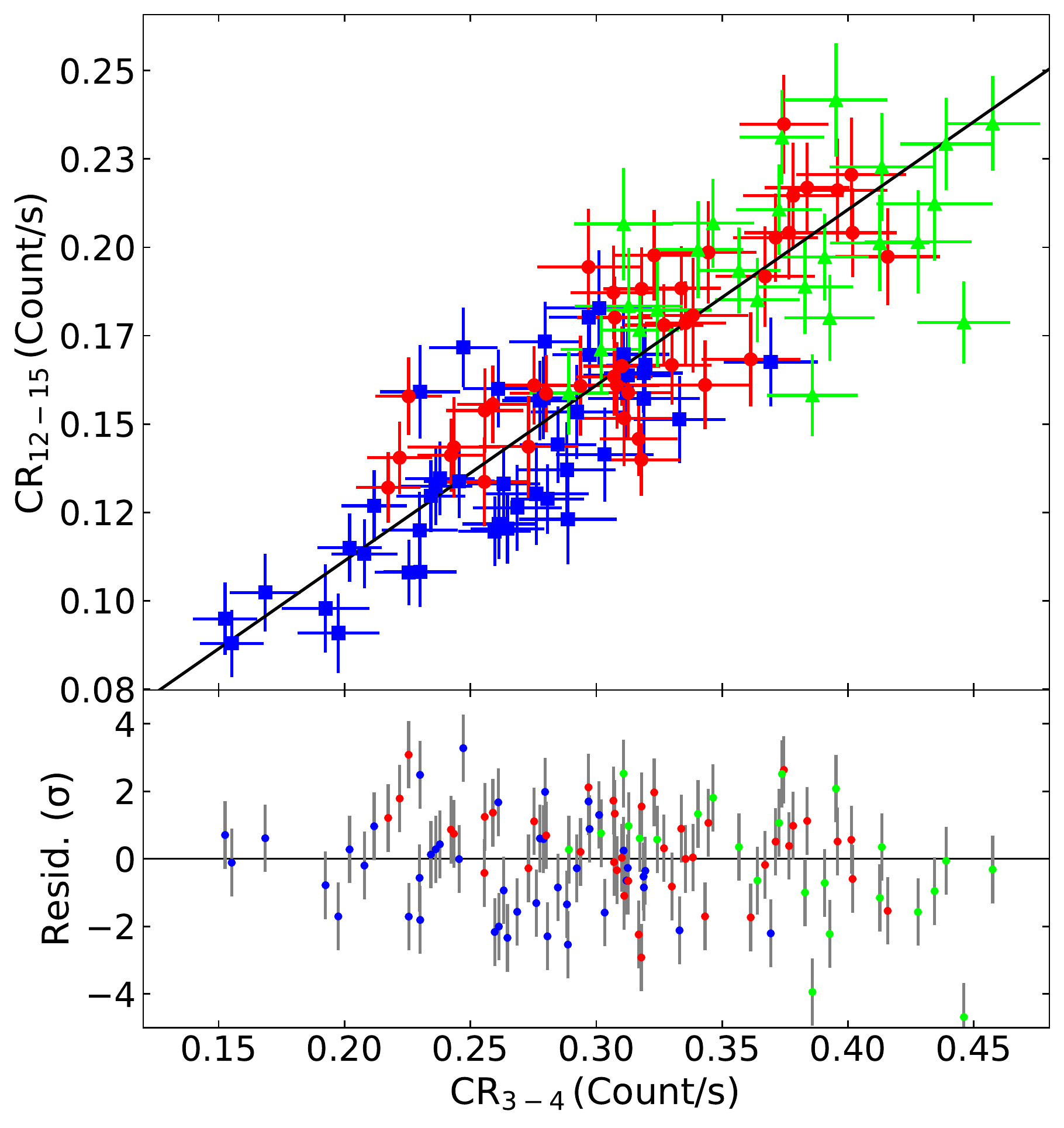}}
{\includegraphics[width=0.235\textwidth]{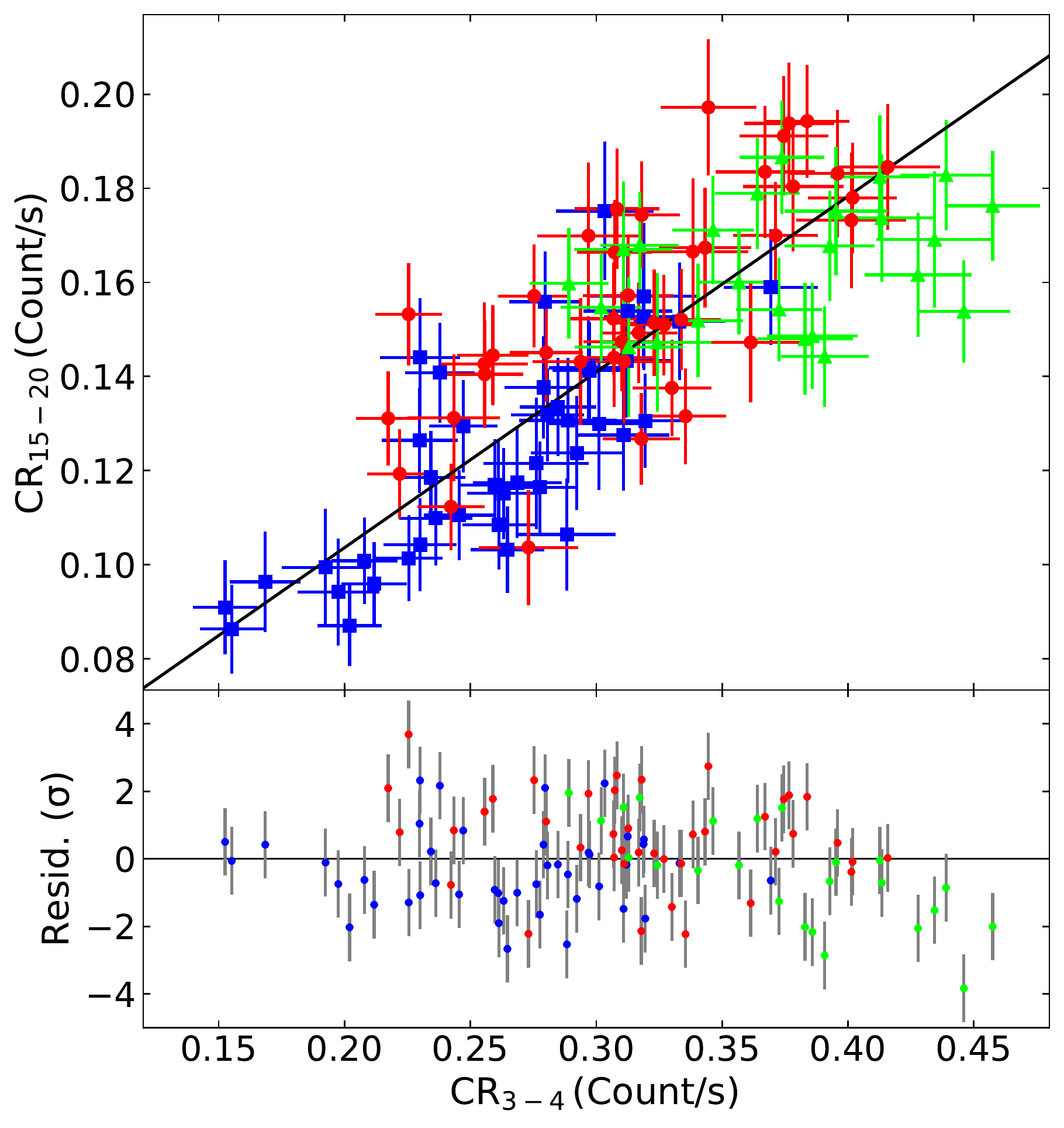}}
{\includegraphics[width=0.235\textwidth]{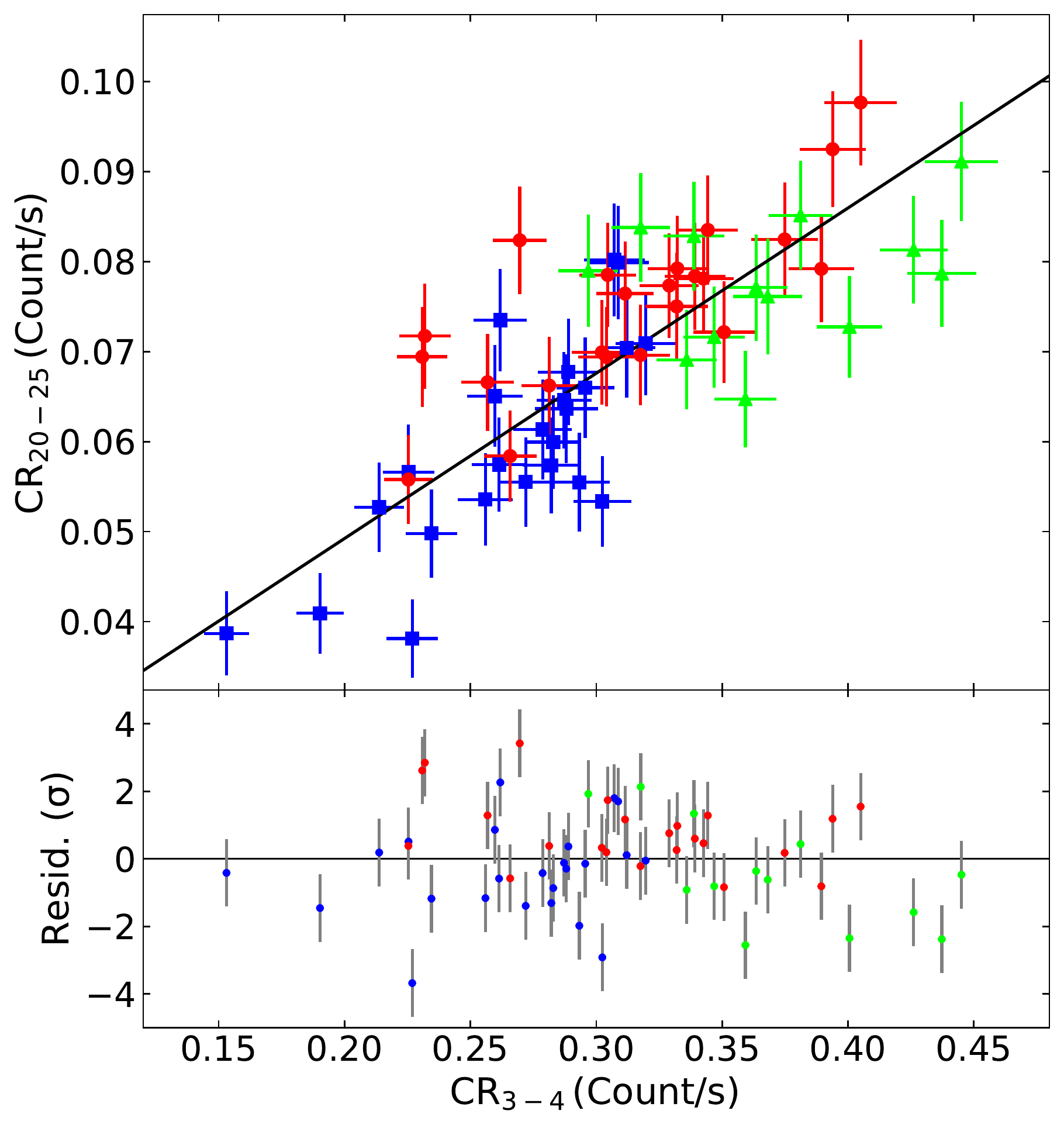}}
{\includegraphics[width=0.235\textwidth]{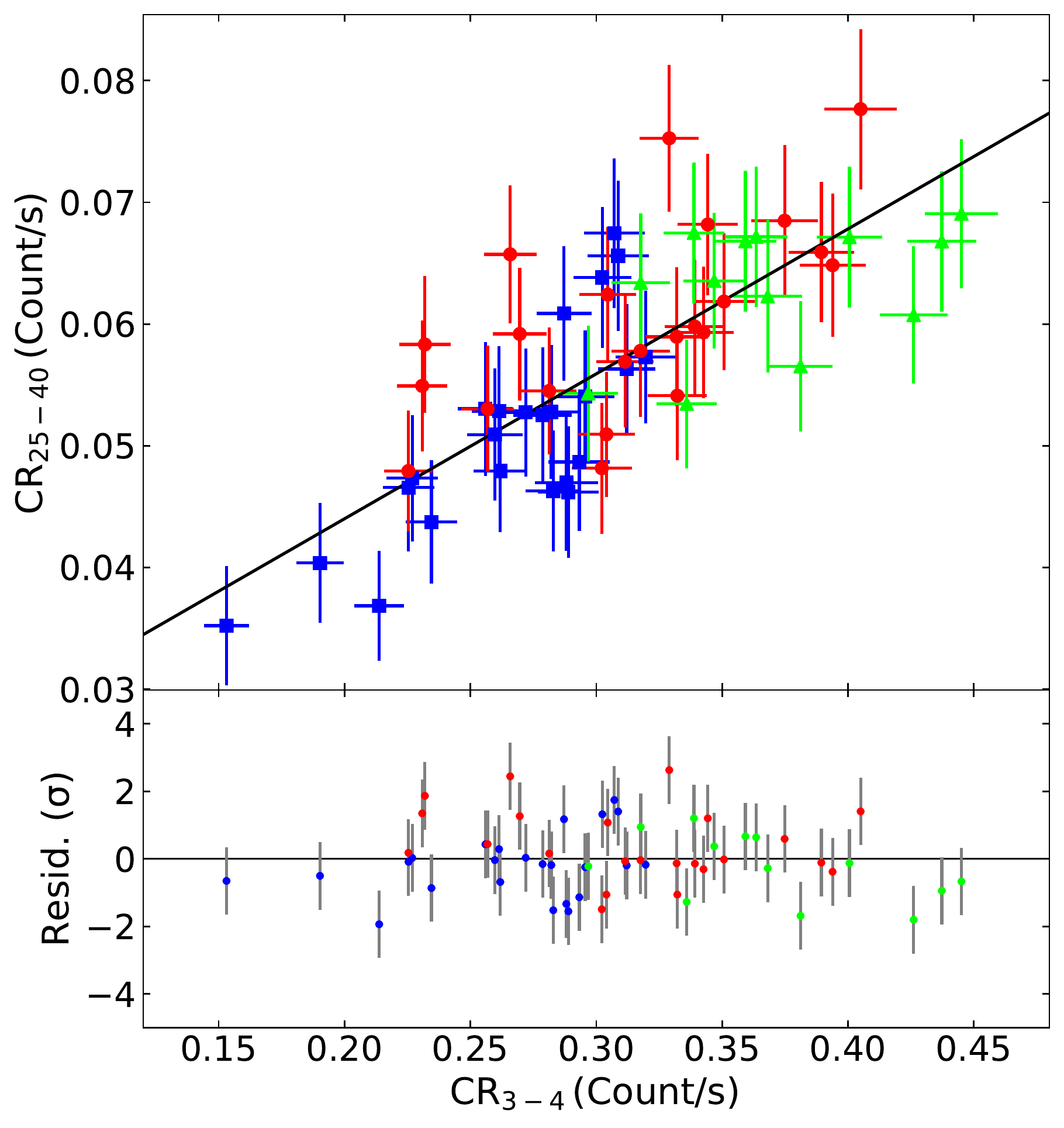}}
\caption{Similar to Fig.\,\ref{figapp:commFFPs} but for the {\it NuSTAR}-only FFPs, in the energy range 10--40\,keV.}
\label{figapp:nustarFFPs}
\end{figure}

\begin{figure}
\centering

{\includegraphics[scale = 0.22]{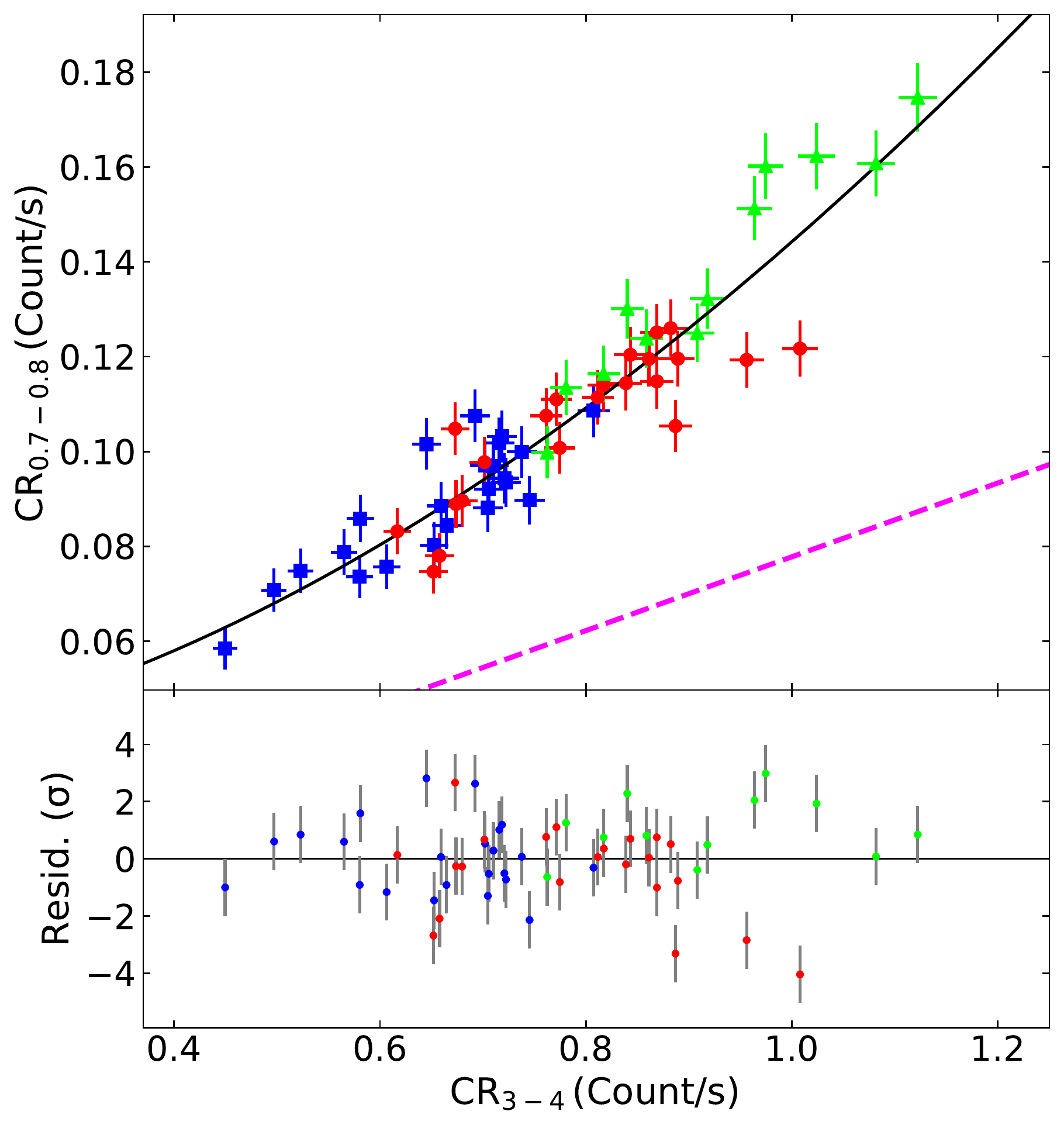}}
{\includegraphics[scale = 0.22]{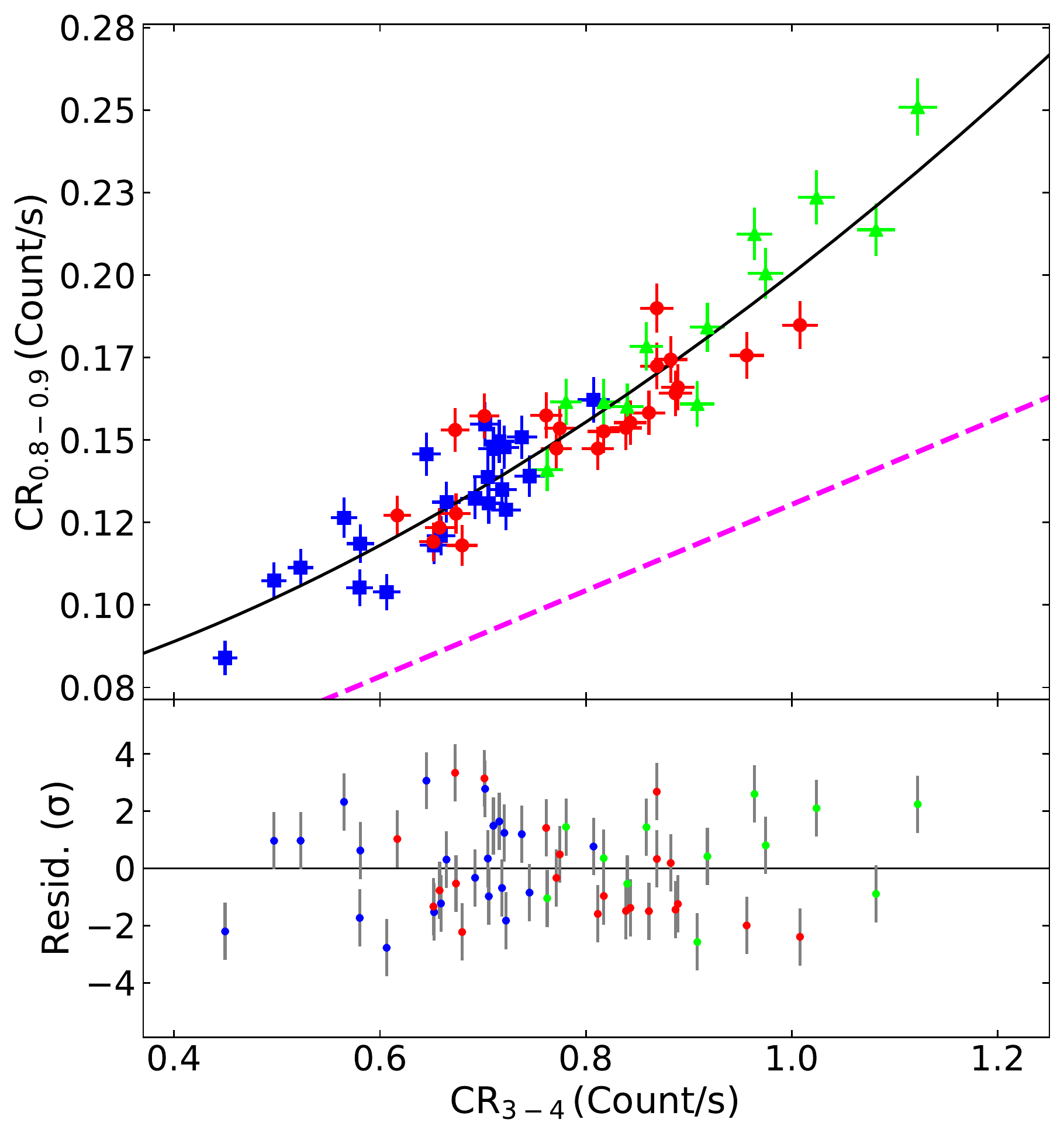}}\\
{\includegraphics[scale = 0.22]{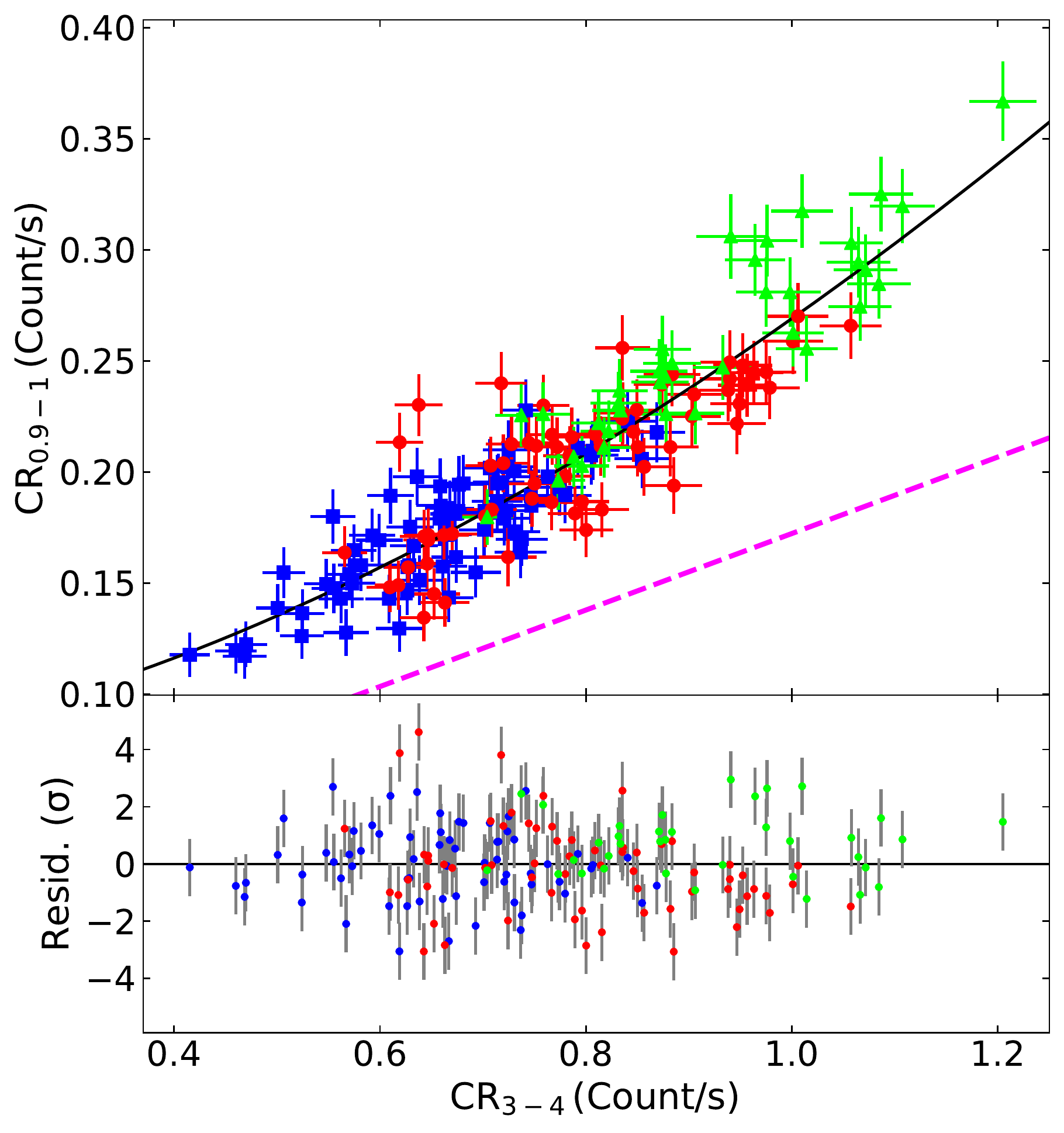}}
{\includegraphics[scale = 0.22]{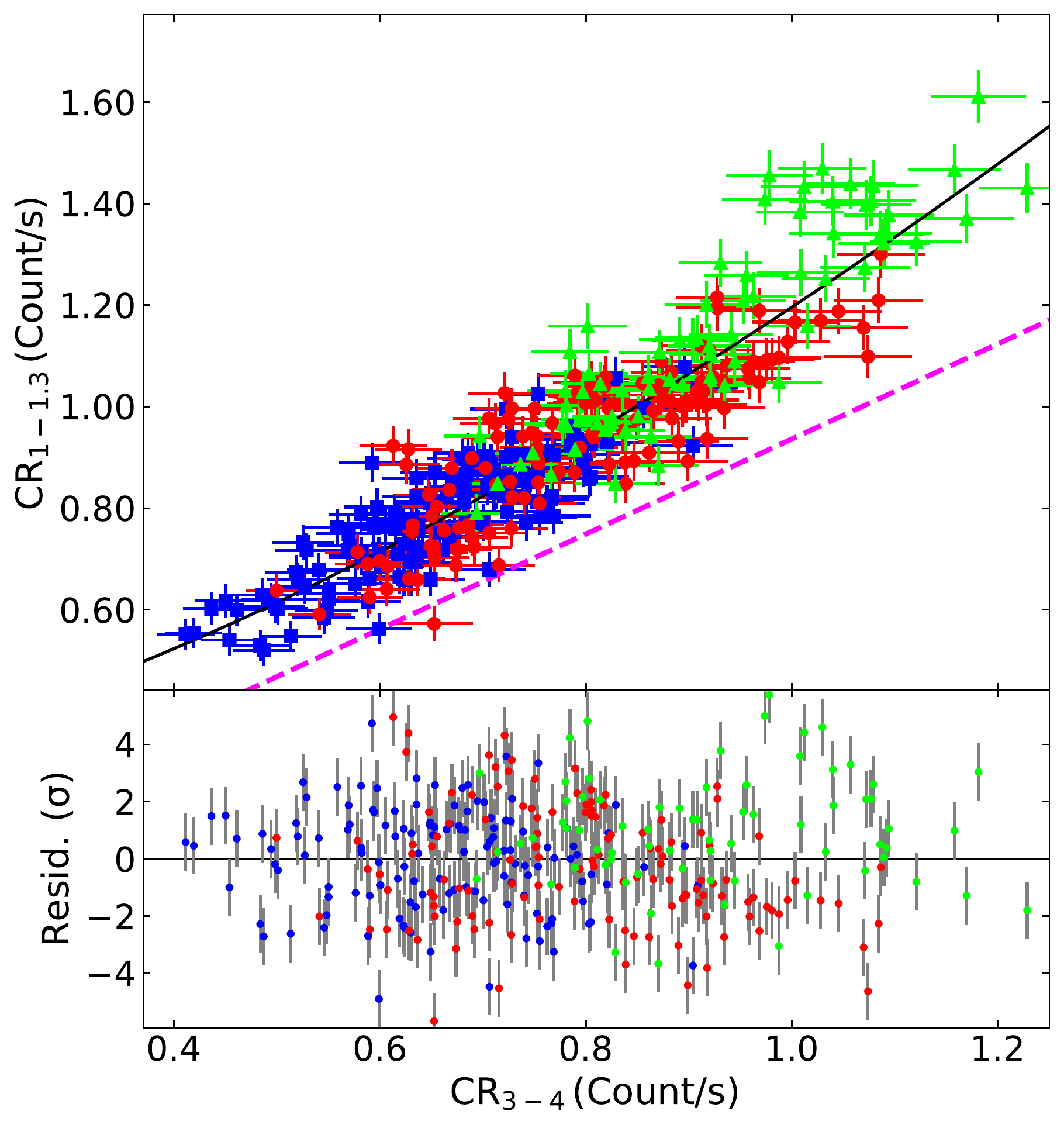}}\\
{\includegraphics[scale = 0.22]{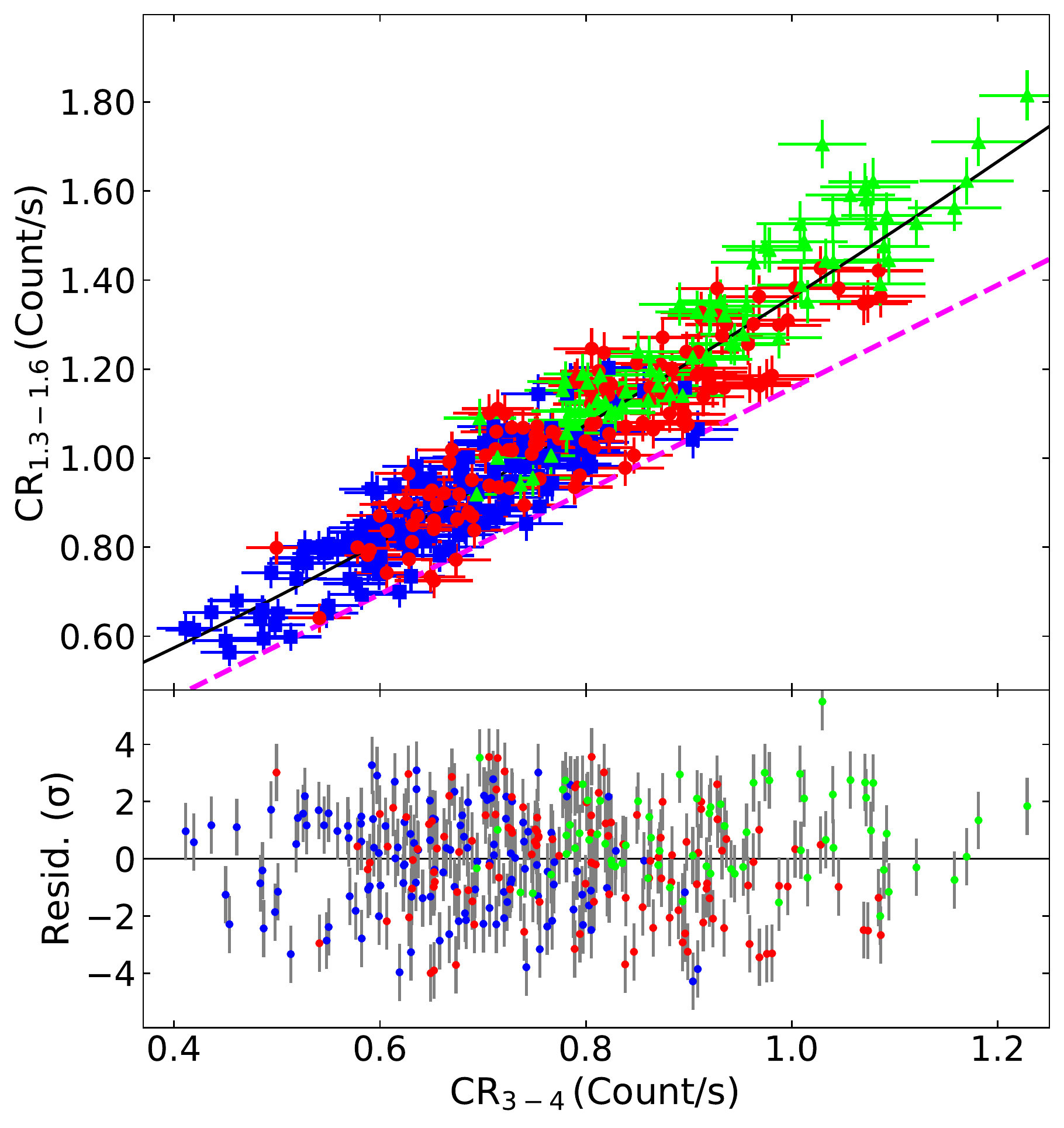}}
{\includegraphics[scale = 0.22]{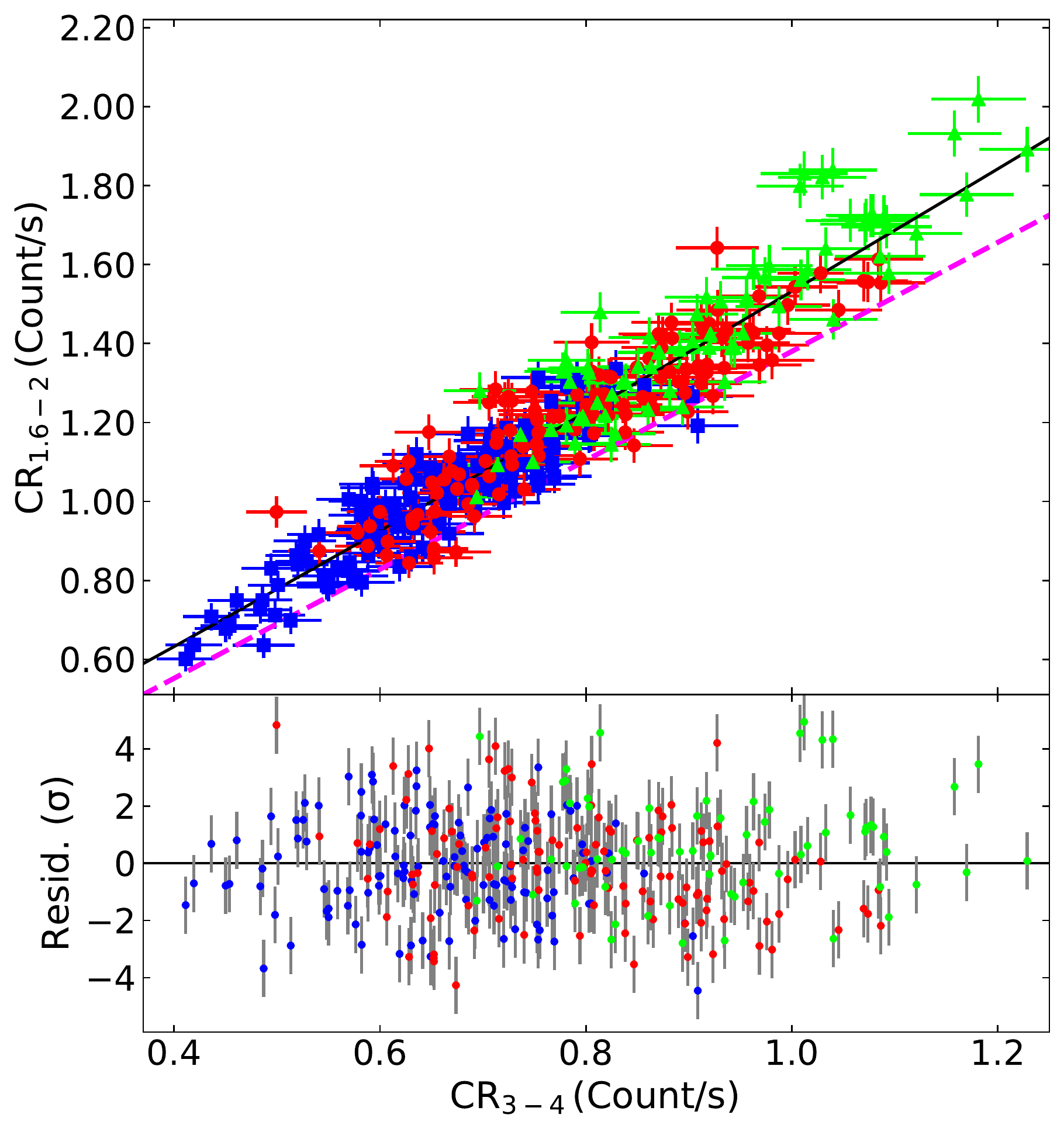}}\\
{\includegraphics[scale = 0.22]{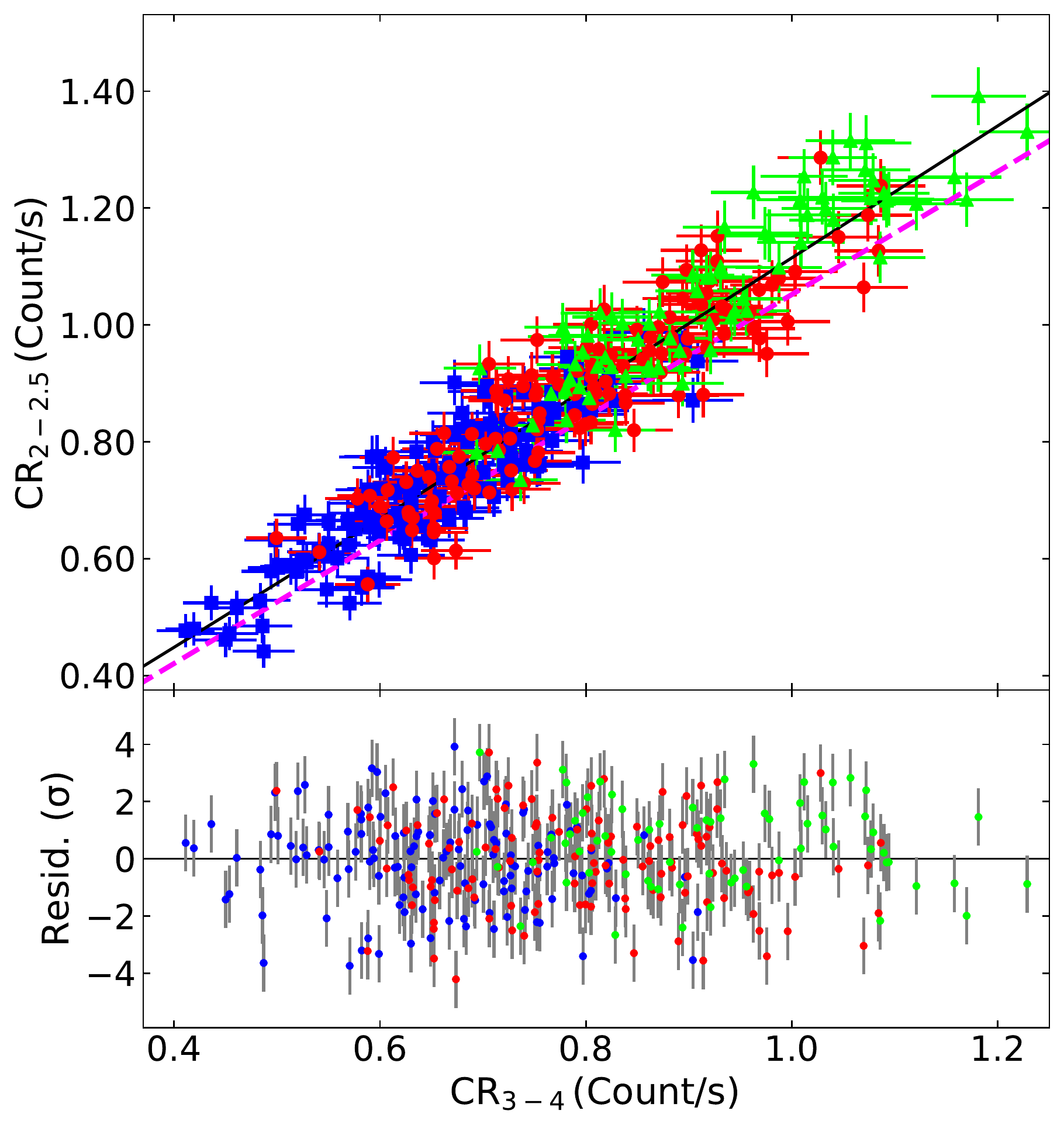}}
{\includegraphics[scale = 0.22]{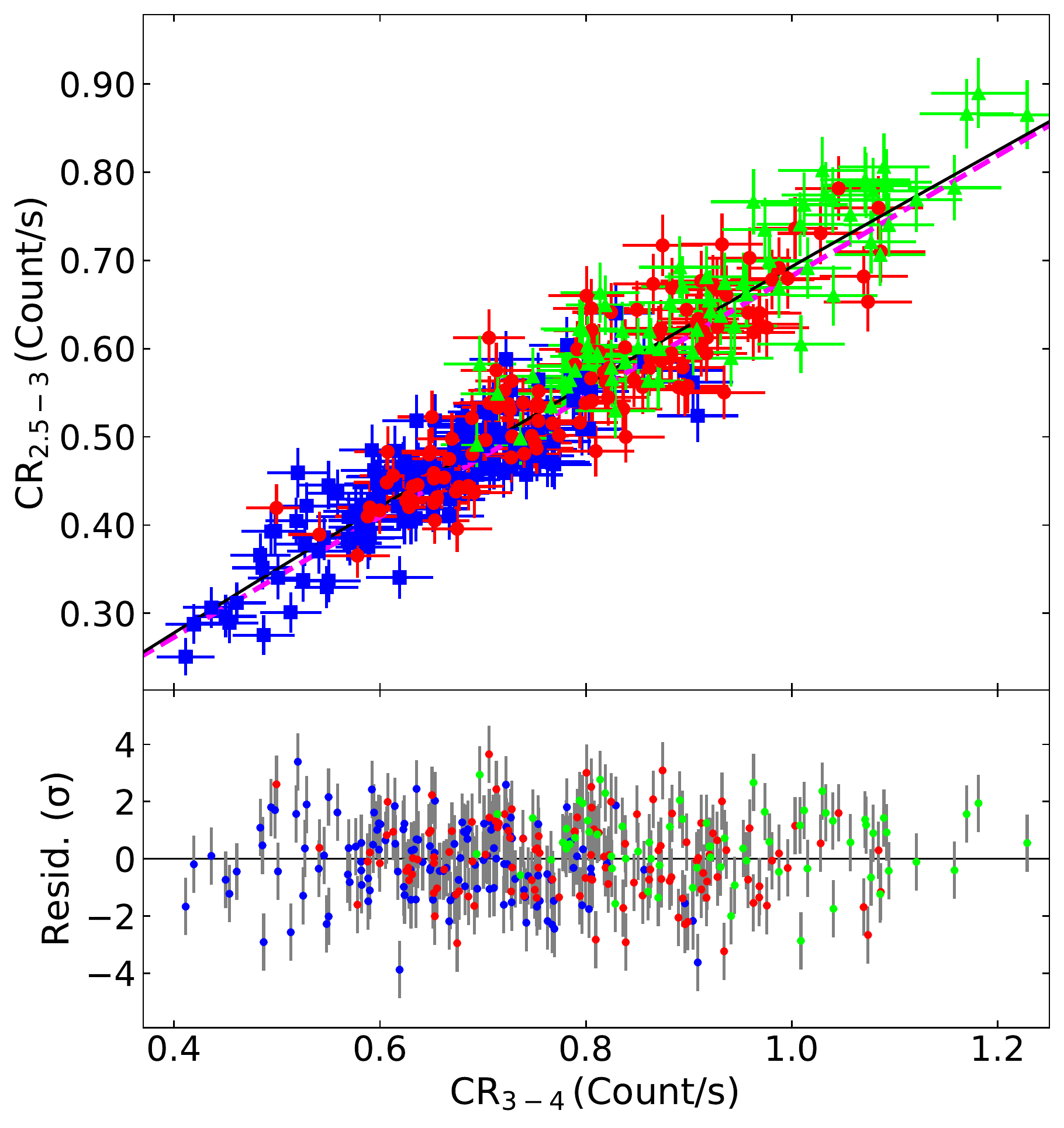}}

\caption{Low-energy FFPs in the 0.7--3\,keV energy range, 
fitting all the observations together. 
The solid lines correspond to the best-fit PLc model obtained 
by fitting the data from each time interval separately, using the 
same color code. The dashed magenta line indicates the {\it predicted} 
FFPs assuming a PL spectrum with $\Gamma_{\rm X} = 1.84$ 
(see Section\,\ref{subsec:highE-FFP} for details). We did not plot the 
error bars for clarity reasons. The best-fit residuals are plotted 
in the lower panel of each plot. }
\label{figapp:lowEFFPs}
\end{figure}

\clearpage
\section{Tables}
\label{app:tables} 

\begin{table}
\centering
\caption{Best-fit parameters obtained by fitting the {\it XMM-Newton} FFPs above 3\,keV with a linear relationship.}
\begin{adjustbox}{max width=1.\linewidth}
\begin{tabular}{lcccc}
\hline	  
Energy Band ($\Delta t$)			&	$	A_{\rm L}			$	&	$	C_{\rm L}			$	&	$	\chi^2/{\rm dof}			$	&	$\sigma_{\rm rms}$	\\[0.1cm]  
[keV (ks)]			&						&	$	({\rm Count/s})			$	&						&		\\    \hline 
4	--	5 (1)	&	$	0.65	\pm	0.01	$	&	$	0.014	\pm	0.010	$	&	$	329	/	332	$	&	0.05	\\[0.1cm]   
5	--	6	(1) &	$	0.45	\pm	0.01	$	&	$	0.014	\pm	0.008	$	&	$	397	/	332	$	&	0.06	\\[0.1cm]   
6	--	7	(1) &	$	0.31	\pm	0.01	$	&	$	0.034	\pm	0.007	$	&	$	364	/	332	$	&	0.05	\\[0.1cm]   
7	--	8	(1.5)&	$	0.17	\pm	0.01	$	&	$	0.016	\pm	0.005	$	&	$	247	/	222	$	&	0.05	\\[0.1cm]   
8	--	10	(2) &	$	0.17	\pm	0.01	$	&	$	0.007	\pm	0.005	$	&	$	198	/	166	$	&	0.04	\\   
  \hline

\end{tabular}
\end{adjustbox}
\label{table:bestfitlinearXMM}
\end{table}

\begin{table}
\centering
\caption{Best-fit parameters obtained by fitting the {\it NuSTAR} FFPs above 3\,keV with a linear relationship.}
\begin{adjustbox}{max width=1.\linewidth}
\begin{tabular}{lcccc}
\hline	  
Energy Band	($\Delta t$)		&	$	A_{\rm L}			$	&	$	C_{\rm L}			$	&	$	\chi^2/{\rm dof}			$	&	$\sigma_{\rm rms}$		\\[0.1cm]  
[keV (ks)]			&						&	$	({\rm Count/s})			$	&						&		\\    \hline 
4	--	5 (1)	&	$	1.04	\pm	0.04	$	&	$	0.009	\pm	0.011	$	&	$	195	/	207	$	&	0.07	\\[0.1cm]   
5	--	6	 (1)	&	$	0.89	\pm	0.03	$	&	$	0.03	\pm	0.010	$	&	$	180	/	207	$	&	0.06	\\[0.1cm]   
6	--	7	 (1)	&	$	0.77	\pm	0.03	$	&	$	0.04	\pm	0.009	$	&	$	179	/	207	$	&	0.05	\\[0.1cm]   
7	--	8	 (1.5)	&	$	0.62	\pm	0.03	$	&	$	0.01	\pm	0.008	$	&	$	121	/	128	$	&	0.06	\\[0.1cm]   
8	--	10	 (1)	&	$	0.94	\pm	0.03	$	&	$	0.02	\pm	0.010	$	&	$	221	/	207	$	&	0.08	\\[0.1cm]   
10	--	12	 (2)	&	$	0.53	\pm	0.02	$	&	$	0.03	\pm	0.007	$	&	$	119	/	108	$	&	0.05	\\[0.1cm]   
12	--	15	 (2)	&	$	0.50	\pm	0.02	$	&	$	0.01	\pm	0.006	$	&	$	161	/	108	$	&	0.08	\\[0.1cm]   
15	--	20	 (2)	&	$	0.37	\pm	0.02	$	&	$	0.03	\pm	0.006	$	&	$	172	/	108	$	&	0.08	\\[0.1cm]   
20	--	25	 (5.8)	&	$	0.18	\pm	0.01	$	&	$	0.01	\pm	0.004	$	&	$	108	/	58	$	&	0.09	\\[0.1cm]   
25	--	40	 (5.8)	&	$	0.12	\pm	0.01	$	&	$	0.02	\pm	0.004	$	&	$	61	/	58	$	&	0.03	\\   
   \hline

\end{tabular}
\end{adjustbox}
\label{table:bestfitlinearNustar}
\end{table}

\begin{table}
\centering
\caption{Best-fit parameters obtained by fitting the {\it XMM-Newton} FFPs below 3\,keV with a PLc relationship.}
\begin{adjustbox}{max width=1.1\linewidth}
\begin{tabular}{lccccc}
\hline	  
Energy	Band ($\Delta t$)	&	$	A_{\rm PLc}			$	&	$	\beta			$	&	$	C_{\rm PLc}			$	&	$	\chi^2/{\rm dof}			$	&	$\sigma_{\rm rms}$	\\[0.1cm]  
[keV (ks)]			&						&	$				$	&	$	({\rm Count/s})			$	&						&		\\   \hline 
0.7	--	0.8	(5.8) &	$	0.11	\pm	0.01	$	&	$	1.77	\pm	0.28	$	&	$	0.037	\pm	0.010	$	&	$	122	/	54	$	&	0.05	\\[0.1cm]   
0.8	--	0.9	(5.8) 	&	$	0.14	\pm	0.01	$	&	$	1.74	\pm	0.26	$	&	$	0.060	\pm	0.013	$	&	$	151	/	54	$	&	0.06	\\[0.1cm]   
0.9	--	1	(2) 	&	$	0.19	\pm	0.01	$	&	$	1.68	\pm	0.20	$	&	$	0.074	\pm	0.015	$	&	$	336	/	165	$	&	0.08	\\[0.1cm]   
1	--	1.3	(1) &	$	0.90	\pm	0.04	$	&	$	1.49	\pm	0.09	$	&	$	0.292	\pm	0.036	$	&	$	1306	/	331	$	&	0.08	\\[0.1cm]   
1.3	--	1.6	(1) 	&	$	1.12	\pm	0.05	$	&	$	1.32	\pm	0.08	$	&	$	0.239	\pm	0.048	$	&	$	1055	/	331	$	&	0.06	\\[0.1cm]   
1.6	--	2	(1) 	&	$	1.44	\pm	0.07	$	&	$	1.07	\pm	0.08	$	&	$	0.093	\pm	0.076	$	&	$	1069	/	331	$	&	0.06	\\[0.1cm]   
2	--	2.5	(1) 	&	$	1.09	\pm	0.07	$	&	$	1.03	\pm	0.09	$	&	$	0.027	\pm	0.069	$	&	$	852	/	331	$	&	0.06	\\[0.1cm]   
2.5	--	3	(1) 	&	$	0.74	\pm	0.07	$	&	$	0.90	\pm	0.11	$	&	$	-0.048	\pm	0.072	$	&	$	582	/	331	$	&	0.05	\\   
\hline

\end{tabular}
\end{adjustbox}
\label{table:bestfitPLcXMM}
\end{table}

\section{Estimation of the variance}
\label{subsec:rms}

The excess variance ($\sigma^2_{\rm XV}$) is commonly used in order to estimate the intrinsic variance of the light curves. We estimated $\sigma^2_{\rm XV}$ of the source as follows.

First we divided the light curve of each energy band into $n=16$ segments of 20~ks long. We estimated $\sigma^2_{\rm XV}$ of each segment as,

\begin{equation}
\sigma^2_{\rm XV} = \frac{1}{N}\sum\limits_{i=1}^N \left[ (y_i - \overline{y})^2 - \sigma_{{\rm err},i}^2 \right],
\end{equation}

\noindent
where $y_i$ and $\sigma_{\rm err,i}$ are the count rate and its error, respectively, $\overline{y}$ is the mean count rate, and $N$ is the number of bins in each segment. We considered the mean excess variance,
\begin{equation}
\overline{\sigma_{\rm XV}^2} = \sum\limits_{i=1}^N \frac{\sigma_{{\rm XV},i}^2}{n}
\end{equation}
and its error,
\begin{equation}\label{eq:meanSig}
{\rm err}\left(\overline{\sigma_{\rm XV}^2}\right) = \sqrt{\frac{\sum\limits_{i=1}^N \left[ \sigma_{{\rm XV},i}^2 - \overline{\sigma_{\rm XV}^2} \right]^2}{n(n-1)}},
\end{equation}
\noindent as the estimate of variance in each energy band. We did not estimate the variance using the whole light curve, because single measurements of $\sigma_{\rm XV}^2$ follow a highly asymmetric distribution, with a large (and unknown) variance. On the other hand, according to \cite{Allevato13} the average estimates which we computed should follow a Gaussian distribution with known errors (the one given by eq.\,\ref{eq:meanSig}). 

\bsp	
\label{lastpage}
\end{document}